\documentclass[times, 10pt]{elsarticle}
\usepackage{comment}
\usepackage{amssymb}
\usepackage{amsmath}
\usepackage{amsthm}
\usepackage{booktabs} 
\newtheorem{theorem}{Theorem}
\newtheorem{lemma}{Lemma}
\usepackage{algorithm}
\usepackage{algorithmic}
\journal{Pattern Recognition}

\begin{document}

\begin{frontmatter}



\title{Implicit Image-to-Image Schrödinger Bridge for Image Restoration}


\author[label1,label2]{Yuang Wang} %% Author name
\author[label2]{Siyeop Yoon}
\author[label2]{Pengfei Jin}
\author[label2]{Matthew Tivnan}
\author[label2]{Sifan Song}
\author[label2]{Zhennong Chen}
\author[label2]{Rui Hu}
\author[label1]{Li Zhang}
\author[label2]{Quanzheng Li}
\author[label1]{Zhiqiang Chen}

\author{Dufan Wu\corref{cor1}\fnref{label2}}
\cortext[cor1]{Corresponding author.}
\ead{dwu6@mgh.harvard.edu}
%{Send correspondence to Dufan Wu (E-mail: dwu6@mgh.harvard.edu}
\affiliation[label1]{organization={The Department of Engineering Physics, Tsinghua University},%Department and Organization
            addressline={30 Shuangqing Road, Haidian}, 
            city={Beijing},
            postcode={100084},
            country={China}}
\affiliation[label2]{organization={Center for Advanced Medical Computing and Analysis, Massachusetts General Hospital and Harvard Medical School},%Department and Organization
            addressline={55 Fruit Street}, 
            city={Boston},
            postcode={MA 02114}, 
            state={Massachusetts},
            country={USA}}

%% Abstract
\begin{abstract}
Diffusion-based models have demonstrated remarkable effectiveness in image restoration tasks; however, their iterative denoising process, which starts from Gaussian noise, often leads to slow inference speeds. The Image-to-Image Schrödinger Bridge (I$^2$SB) offers a promising alternative by initializing the generative process from corrupted images while leveraging training techniques from score-based diffusion models. In this paper, we introduce the Implicit Image-to-Image Schrödinger Bridge (I$^3$SB) to further accelerate the generative process of I$^2$SB. I$^3$SB restructures the generative process into a non-Markovian framework by incorporating the initial corrupted image at each generative step, effectively preserving and utilizing its information. To enable direct use of pretrained I$^2$SB models without additional training, we ensure consistency in marginal distributions. Extensive experiments across many image corruptions—including noise, low resolution, JPEG compression, and sparse sampling—and multiple image modalities—such as natural, human face, and medical images— demonstrate the acceleration benefits of I$^3$SB. Compared to I$^2$SB, I$^3$SB achieves the same perceptual quality with fewer generative steps, while maintaining or improving fidelity to the ground truth.
\end{abstract}



%% Keywords

\end{frontmatter}



\section{Introduction}
Restoring high-quality images from degraded ones is a fundamental yet challenging task in both natural and medical imaging. Image corruptions, can arise from various factors, including noise,  low resolution, compression artifacts and sparse sampling.  Recently, conditional diffusion models~\cite{dhariwal2021diffusion,saharia2022image} have shown promising performance in addressing this challenge. Rooted in stochastic process theories, diffusion models offer a more stable approach to sampling from complex distributions compared to Generative Adversarial Networks (GANs)~\cite{gonzalez2024dgd}. However, the inference speed of diffusion models is often limited by the large number of iterative denoising steps needed to generate clean images starting from pure Gaussian noise. 

Instead of starting from Gaussian noise, Schrödinger Bridges establish diffusion bridges between the distributions of clean and corrupted images. By initiating the diffusion process with the corrupted image, which is closer to the clean one than Gaussian noise, Schrödinger Bridges offer a promising approach to generate high-quality conditional samples with fewer diffusion steps. A notable instance of this framework is the Image-to-Image Schrödinger Bridge (I$^2$SB)\cite{liu20232}, which models the transition between paired clean and corrupted images, facilitating efficient training through its connection to score-based diffusion models. The number of neural function evaluations (NFEs) in its generative process controls the trade-off between perceptual quality and fidelity to the ground truth~\cite{chung2024direct}. With a small NFE, I$^2$SB tends to produce less distortion but smoother images due to the ``regression to the mean effect"~\cite{delbracio2023inversion}, whereas with a large NFE, I$^2$SB can generate images with high perceptual quality at the expense of some distortion from the ground truth.

In this work, our aim is to accelerate I$^2$SB to achieve the same perceptual quality and equal or better fidelity to the ground truth with fewer generative steps. Inspired by the success of incorporating non-Markovian processes into the denoising diffusion probabilistic model (DDPM)~\cite{ho2020denoising} to create the denoising diffusion implicit model (DDIM)~\cite{song2020denoising}, we proposed the Implicit Image-to-Image Schrödinger Bridge (I$^3$SB). Our approach employs a non-Markovian process during inference by incorporating the initial corrupted image into each step alongside the current image and the estimated mean, effectively preserving and utilizing the information from corrupted images. By constraining the marginal distribution to match that of I$^2$SB, the proposed I$^3$SB shares the same training loss functions with I$^2$SB and can thus reuse pretrained I$^2$SB models. We introduced a single parameter to balance the Markovian and non-Markovian components. Additionally, we established a connection between I$^3$SB and the probability flow ordinary differential equation (PF-ODE) of the Variance Exploded (VE) endpoint-fixed Schrödinger Bridge~\cite{zhou2023denoising}.

We validate I$^3$SB in many  image restoration tasks on multiple image modalities, including super-resolution and JPEG restoration for natural and human face images, as well as CT sparse-view reconstruction, super-resolution, and denoising for medical images. Extensive experiments demonstrate that, compared to I$^2$SB, I$^3$SB achieves similar perceptual quality with fewer NFEs while maintaining equal or superior fidelity to the ground truth. Additionally, I$^3$SB outperforms several state-of-the-art diffusion-based image restoration models in both quantitative metrics and visual quality.

The main contributions of our work can be summarized as follows:
\begin{itemize}
\item{We introduce I$^3$SB, an innovative framework that modifies the generative process of I$^2$SB from a Markovian to a non-Markovian approach, incorporating the corrupted image at each generative step, effectively reducing information loss.}
\item{We establish a theoretical connection between I$^3$SB and the PF-ODE of the VE endpoint-fixed Schrödinger Bridge.}
\item{Through extensive experiments on image corruptions, including noise, low resolution, JPEG compression, and sparse sampling, we demonstrate the accelerated performance of I$^3$SB. Compared to I$^2$SB, I$^3$SB achieves similar perceptual quality with fewer NFEs while maintaining or improving fidelity to the ground truth.}
\end{itemize}

\label{sec1}
%% Labels are used to cross-reference an item using \ref command.
\section{Related Work}
\subsection{Image Restoration Methods}
Image restoration is a critical task in image processing and pattern recognition, focused on recovering high-quality images from degraded inputs, such as those affected by noise, low resolution, compression artifacts and sparse sampling. Traditional methods typically rely on predefined priors and optimization techniques for image reconstruction. Filtering-based approaches, such as Gaussian smoothing and Non-Local Means Filtering~\cite{buades2011non}, work by reducing noise while preserving edges through averaging nearby or similar pixels. Variational methods~\cite{el2010weighted,li2022joint} assume that images have sparse representations in specific transformation domains (e.g., wavelet or Fourier), framing image restoration as an optimization problem that minimizes a loss function containing both a data consistency term and a regularization term, such as total variation regularization. Dictionary learning approaches model images as sparse combinations of atoms from a learned dictionary, using methods like K-SVD~\cite{aharon2006k} to build dictionaries and restore clean images.

In recent years, deep learning has revolutionized image restoration by replacing handcrafted priors with data-driven representations, enabling models to learn complex image patterns directly from data. Convolutional Neural Networks (CNNs) learn end-to-end mappings from corrupted to clean images. While these methods excel in fidelity, they often produce overly smooth images. Vision Transformers (ViTs), like Restormer~\cite{zamir2022restormer}, leverage self-attention mechanisms to capture long-range dependencies in images, leading to significant improvements in detail restoration. Conditional Variational Autoencoders (CVAEs)~\cite{harvey2022conditional} learn compact latent representations of clean images, providing a probabilistic framework for generating clean images from Gaussian noise conditioned on the corrupted inputs. Normalizing flow-based models model complex data distributions through invertible transformations, serving as priors to regularize the image restoration process~\cite{helminger2021generic}. GAN-based models, such as DGD-cGAN~\cite{gonzalez2024dgd}, use adversarial training to generate visually realistic textures, although they may suffer from unstable training and issues like mode collapse. Recently, score-based diffusion models have emerged as state-of-the-art methods for image restoration~\cite{liu2024adaptbir}, and we will provide an overview of these models in the following subsection.

\subsection{Diffusion Based Image Restoration Models}
Diffusion based image restoration models can be broadly categorized into task-agnostic and task-specific models. Task-agnostic models train unconditional score functions with denoising score matching~\cite{song2020score} as priors for the data distribution, and incorporate data consistency during inference. These models can be used for various types of corruptions with the same trained network, but they require the specific forward operator corresponding to each corruption during inference. For instance, DDNM~\cite{wang2022zero} and DDRM~\cite{kawar2022denoising,kawar2022jpeg} decompose images into the range and null spaces of the forward operator, applying data consistency to the range space component.  DPS~\cite{chung2022diffusion} and $\Pi$GDM~\cite{song2023pseudoinverse} sample from the posterior distribution, using approximations to make the process tractable. Red-Diff~\cite{mardani2023variational} formulates image restoration as an optimization problem, incorporating the pretrained score function as regularization.

Task-specific models train conditional score functions using the corrupted image as a condition. These models require separate training for each type of corruption, necessitating pairs of clean and corrupted images during training. However, they often outperform task-agnostic models. Moreover, because task-specific models do not need to incorporate additional data consistency during inference, they are capable of handling blind corruption scenarios where the forward operator is unknown. Notable examples of task-specific models include SR3~\cite{saharia2022image} and ADM~\cite{dhariwal2021diffusion}.

\subsection{Acceleration for Diffusion Models}
Accelerating the inference process of diffusion models has become a key focus in the field, leading to the development of two main approaches. The first approach involves generating samples by solving the PF-ODE using high-order solvers instead of sampling from stochastic differential equations (SDEs). These methods modify only the inference process without further training the score function. For instance, DDIM~\cite{song2020denoising} accelerates DDPM by transforming the Markovian process into a non-Markovian process and establishes its connection with the PF-ODE. EDM~\cite{karras2022elucidating}  solve the PF-ODE using the Heun's 2$^{\text{nd}}$ order method, achieving state-of-the-art results in image generation. DPM-solver~\cite{lu2022dpm} transforms the ODE discretization into a discretization of the exponentially weighted integral of the score function, applying Taylor expansion for approximation. PNDM~\cite{liu2022pseudo} introduces a manifold-constrained high-order ODE solver, utilizing linear multi-step and Runge-Kutta methods while maintaining intermediate images on the true noisy manifold.

Diffusion distillation is another approach that accelerates the diffusion inference process by leveraging the PF-ODE, which establishes a deterministic mapping between Gaussian noise and the final generated images.  These methods train student models to distill the multi-step outputs of the original diffusion model into a single step. For example, Progressive Distillation~\cite{salimans2022progressive} introduces binary time distillation, where the student model is trained to predict the two-step output of the teacher model and then serves as the teacher in the next phase. Consistency Model~\cite{song2023consistencymodels} uses the student model, equipped with exponentially moving averaged weights, as a self-teacher to incorporate self-consistency into the student model.

\subsection{Paired Data Schrödinger Bridge}
Unlike diffusion models that start from Gaussian noise, Schrödinger Bridges initiate their generative processes from corrupted images, offering a promising alternative for generating high-quality conditional samples with fewer generative steps. Schrödinger Bridges can be broadly categorized into unpaired and paired data Schrödinger Bridges. While unpaired data Schrödinger Bridges~\cite{chen2022likelihood} often face challenges with inefficient training, paired data Schrödinger Bridges, such as I$^2$SB~\cite{liu20232} and InDI~\cite{delbracio2023inversion}, can be trained as efficiently as score-based diffusion models. Techniques developed for diffusion models have been adapted to paired data Schrödinger Bridges. For instance, DDBM~\cite{zhou2023denoising} applies Heun's 2nd-order method to solve the PF-ODE of paired data Schrödinger Bridges and demonstrates good performance in image translation tasks. CDDB~\cite{chung2024direct} enhances I$^2$SB by incorporating data consistency with techniques from DDS~\cite{chung2023decomposed} and DPS. Our proposed I$^3$SB, inspired by DDIM and transforming the generative process from Markovian to non-Markovian, demonstrates acceleration benefits compared to I$^2$SB.

\section{Preliminaries}
Notation: Let $X_t \in \mathbb{R}^d$ represent a $d$-dimensional stochastic process indexed by $t \in \left[0,1\right]$, and $N$ denote the number of generative steps. We denote the discrete generative time steps as $0=t_0< \cdots t_n \cdots <t_N=1$, and shorthand $X_n \equiv X_{t_n}$.
%% Use \subsection commands to start a subsection.
\subsection{Image-to-Image Schrödinger Bridge}
\label{subsec1}
The I$^2$SB~\cite{liu20232} establishes direct diffusion bridges between paired clean and corrupted images. With $X_0$ representing clean images and $X_1$ representing corresponding corrupted images, $X_t$ is designed to follow the Gaussian distribution $q\left(X_t|X_0,X_1\right)$:
\begin{equation}
  q\left(X_t|X_0,X_1\right)=
  \mathcal N\left(X_t;\frac{\overline{\sigma}_t^2}{\sigma_1^2}X_0+\frac{{\sigma}_t^2}{\sigma_1^2}X_1,\frac{{\sigma}_t^2\overline{\sigma}_t^2}{\sigma_1^2}I\right),  
  \label{eq:mu_t}
\end{equation}
where ${\sigma}_t^2=\int_0^t\beta_\tau d\tau$ and $\overline{\sigma}_t^2=\int_t^1\beta_\tau d\tau$ represent variances accumulated from either side, ${\sigma}_1^2=\int_0^1\beta_\tau d\tau$, and $\beta_\tau$ determines the speed of diffusion. Since $X_t$ can be sampled analytically using equation (\ref{eq:mu_t}), the network $\epsilon_\theta$ can be efficiently trained to predict the difference between $X_t$ and $X_0$ by minimizing the loss function:
\begin{equation}
\theta^*=\arg\min_{\theta}\mathbb{E}_{X_0, X_1}\mathbb{E}_{t,X_t}\Vert \epsilon_{\theta}\left(X_t,t\right)-\frac{X_t-X_0}{\sigma_t} \Vert,
\label{eq:loss}
\end{equation}
where $t \sim \mathcal{U}[0,1]$ and $X_t\sim q\left(X_t|X_0,X_1\right)$.

In the generative process, I$^2$SB begins with the corrupted image $X_N$ and iteratively approaches the clean image $X_0$. In the step from $X_n$ to $X_{n-1}$, $\hat{X}_0^{(n)}$, the expected mean of $X_0$ at time $t_n$, is first calculated using the trained network $\epsilon_{\theta^*}$ and $X_n$:

\begin{equation}
\hat{X}_0^{(n)} = X_{n}-\sigma_{n}\epsilon_{\theta^*}\left(X_{n},t_{n}\right),
\label{eq:x0_hat}
\end{equation}
where we use $\sigma_n \equiv \sigma_{t_n}$. Subsequently, $X_{n-1}$ is sampled from the DDPM posterior $p$ described by $\hat{X}_0^{(n)}$ and $X_{n}$:
\begin{equation}
X_{n-1} \sim p\left(X_{n-1}|\hat{X}_0^{(n)}, X_{n}\right).
\label{eq:i2sb_xn}
\end{equation}
Here, the DDPM posterior $p$ is expressed as:
\begin{equation}
p\left(X_{n-1}|\hat{X}_0^{(n)}, X_{n}\right)=
\mathcal{N}\left(X_{n-1};\frac{\alpha_{n-1}^2}{\sigma_{n}^2}\hat{X}_0^{(n)}+\frac{\sigma_{n-1}^2}{\sigma_{n}^2}X_{n},\frac{\sigma_{n-1}^2\alpha_{n-1}^2}{\sigma_{n}^2}I\right),
\label{eq:p}
\end{equation}
where $\alpha_{n-1}^2=\int_{t_{n-1}}^{t_{n}}\beta_\tau d\tau$ denotes the accumulated variance between consecutive time steps $t_{n-1}$ and $t_n$.
\section{Method}
\subsection{Implicit Image-to-Image Schrödinger Bridge}
\begin{figure*}[!t]
\centering
\includegraphics[width=\textwidth]{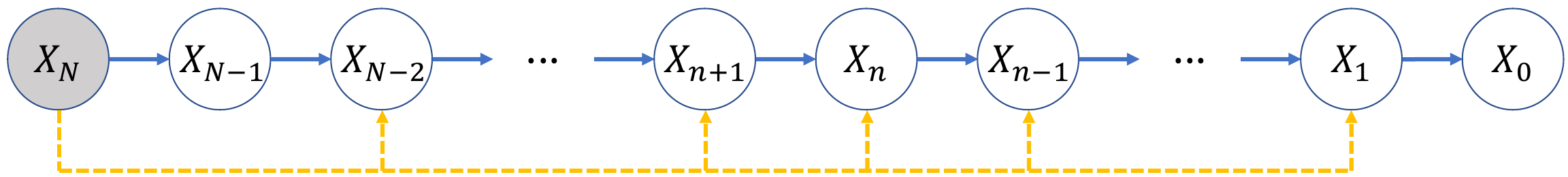}
\caption{Non-Markovian generative process of I$^3$SB. Solid arrows denote original dependencies in I$^2$SB, and dotted arrows signify additional dependencies in I$^3$SB.}
\label{fig1}
\end{figure*}
%% Use \subsubsection, \paragraph, \subparagraph commands to 
%% start 3rd, 4th and 5th level sections.
%% Refer following link for more details.
%% https://en.wikibooks.org/wiki/LaTeX/Document_Structure#Sectioning_commands
\begin{algorithm}
   \renewcommand{\algorithmicrequire}{\textbf{Input:}}
    \caption{Generative Process of I$^3$SB}
    \label{alg1}
    \begin{algorithmic}
		\REQUIRE $N$, $\left\{t_n\right\}$, corrupted image $X_N$, trained network $\epsilon_{\theta^*}$
		\FOR{$n=N$ to $1$} 
		\STATE Predict $\hat{X}_0^{(n)}$ using $\epsilon_{\theta^*}\left(X_n, t_n\right)$ and $X_n$
		\IF{$n == N$} 
		\STATE Sample $X_{n-1}$ from $p\left(X_{n-1}|\hat{X}_0^{(n)}, X_n\right)$
		\ELSIF{$1<n<N$}
		\STATE Sample $X_{n-1}$ from $p_G\left(X_{n-1}|\hat{X}_0^{(n)}, X_n,X_N\right)$
            \ELSE
            \STATE $X_0=\hat{X}_0^{(n)}$
		\ENDIF 
            \ENDFOR
        \RETURN $X_0$
    \end{algorithmic}
\end{algorithm}
\subsubsection{Motivation}
We preserve the training process of I$^2$SB and aim to accelerate its generative process in our proposed I$^3$SB. The generative process of I$^2$SB is essentially a Markovian chain, where $X_{n-1}$ depends solely on $X_n$ given the trained network $\epsilon_{\theta^*}$ (with $\hat{X}_0^{(n)}$ also being a function of $X_{n}$). In this setup, the information in $X_N$ is only used in the first step and may be gradually lost through the Markovian chain. I$^3$SB fully utilizes the information in $X_N$ by incorporating $X_N$ in each generative step, changing the generative process to a non-Markovian chain, as shown in Figure \ref{fig1}. 

\subsubsection{Algorithm}
In the generative step from $X_n$ to $X_{n-1}$, we first compute $\hat{X}_0^{(n)}$ using equation (\ref{eq:x0_hat}) and then sample $X_{n-1}$ from a distribution $p_G$:
\begin{equation}
X_{n-1} \sim p_G\left(X_{n-1}|\hat{X}_0^{(n)}, X_{n},X_N\right),
\label{eq:i3sb_xn}
\end{equation}
where $X_N$ is included. Following DDIM~\cite{song2020denoising}, we design $p_G\left(X_{n-1}|\hat{X}_0^{(n)},X_{n},X_N\right)$ as a Gaussian distribution with a linear combination of $\hat{X}_0^{(n)}$, $X_{n}$ and $X_N$ as its mean and $g_n^2I$ as its covariance matrix :
\begin{equation}
p_G\left(X_{n-1}|\hat{X}_0^{(n)},X_{n},X_N\right)=\\\mathcal{N}\left(X_{n-1};A_n\hat{X}_0^{(n)}+B_nX_{n}+C_nX_N,g_n^2I\right),
\label{eq:p_G}
\end{equation}
where $A_n$, $B_n$ and $C_n$ represent the weights of $\hat{X}_0^{(n)}$, $X_n$ and $X_N$, respectively, and $g_n$ is a hyperparameter to be discussed later. Since the trained network from I$^2$SB is directly used to predict $\hat{X}_0^{(n)}$ in equation (\ref{eq:x0_hat}), the marginal distribution of I$^3$SB should align with that of I$^2$SB. Specifically, if $\hat{X}_0^{(n)}$ equals to $X_0$ for any $n$, $X_{n-1}$ sampled from $p_G\left(X_{n-1}|X_0,X_n,X_N\right)$ should follow the distribution of $q\left(X_{n-1}|X_0,X_N\right)$. Therefore, the distribution $p_G$ must satisfy the following equation: 
\begin{equation}
q\left(X_{n-1}|X_0,X_N\right)=\int p_G\left(X_{n-1}|X_0,X_{n},X_N\right)q\left(X_{n}|X_0,X_N\right)\text{d}X_{n}.
\label{eq:i3sb_sample}
\end{equation}

Substituting equation (\ref{eq:p_G}) into equation ({\ref{eq:i3sb_sample}}) with $\hat{X}_0^{(n)}$ equal to $X_0$, the weights $A_n$, $B_n$ and $C_n$ can be analytically expressed in terms of $g_n$:
\begin{subequations}\label{eq:ABC}
\begin{align}
A_n&=\frac{\overline{\sigma}_{n-1}^2}{\sigma_N^2}-\frac{\overline{\sigma}_{n}^2}{\sigma_{N}^2}\frac{\sqrt{\sigma_{n-1}^2\overline{\sigma}_{n-1}^2-g_n^2\sigma_N^2}}{\sigma_{n}\overline{\sigma}_{n}},\\
B_n &= \frac{\sqrt{\sigma_{n-1}^2\overline{\sigma}_{n-1}^2-g_n^2\sigma_N^2}}{\sigma_n\overline{\sigma}_n},\\
C_n&=\frac{\sigma_{n-1}^2}{\sigma_N^2}-\frac{\sigma_n^2}{\sigma_{N}^2}\frac{\sqrt{\sigma_{n-1}^2\overline{\sigma}_{n-1}^2-g_n^2\sigma_N^2}}{\sigma_{n}\overline{\sigma}_{n}}.
\end{align}
\end{subequations}
Therefore, $X_{n-1}$ can be efficiently sampled from $p_G \left(X_{n-1}|\hat{X}_0^{(n)}, X_{n},X_N\right)$ using:
\begin{equation}
X_{n-1}=\left(\frac{\overline{\sigma}_{n-1}^2}{\sigma_N^2}\hat{X}_0^{(n)}+\frac{\sigma_{n-1}^2}{\sigma_N^2}X_N\right)+\sqrt{\frac{\sigma_{n-1}^2\overline{\sigma}_{n-1}^2}{\sigma_N^2}-g_n^2}\hat{\epsilon}^{(n)}+g_n\epsilon,
\label{eq:i3sb_xn_2}
\end{equation}
where $\epsilon\sim\mathcal{N}\left(0,I\right)$ is Gaussian noise, and $\hat{\epsilon}^{(n)}$ is the normalized estimated Gaussian noise from $X_{n}$, defined as:
\begin{equation}
\hat{\epsilon}^{(n)}=\frac{\sigma_N}{\sigma_{n}\overline{\sigma}_{n}}\left(X_{n}-\left(\frac{\overline{\sigma}_{n}^2}{\sigma_N^2}\hat{X}_0^{(n)}+\frac{\sigma_{n}^2}{\sigma_N^2}X_N\right)\right).
\label{eq:epsilon_hat}
\end{equation}
The proofs for equations (\ref{eq:ABC}) and (\ref{eq:i3sb_xn_2}) are provided in \ref{appA1}. The generative process of I$^3$SB is summarized in Algorithm \ref{alg1}.

\subsubsection{Hyperparameter}
The term $g_n$ can be freely designed as long as it adheres to the constraint:
\begin{equation}
0\leq g_n\leq\frac{\sigma_{n-1}\overline{\sigma}_{n-1}}{\sigma_N},
\label{eq:gn_constraint}
\end{equation}
to ensure the meaningfulness of taking the square root in equation (\ref{eq:i3sb_xn_2}). The functionality of $g_n$ can be interpreted from multiple aspects: it balances between the Markovian and non-Markovian components, controls the stochasticity of the generative process, and determines the degree of dependence on the deterministic estimate of noise component. When $g_n$ equals to $\frac{\sigma_{n-1}\alpha_{n-1}}{\sigma_{n}}$ for any $n$, $p_G$ becomes equivalent to the DDPM posterior $p$, and the generative process of I$^3$SB reverts to that of I$^2$SB. We parameterized $g_n$ as:
\begin{equation}
g_n =\eta\frac{\sigma_{n-1}\alpha_{n-1}}{\sigma_n},
\label{eq:gn}
\end{equation}
where $\eta$ is left as a hyperparameter.
\subsubsection{Relevance to PF-ODE}
When $g_n$ is set to $0$, the generative process becomes fully deterministic. We provide the ODE for the deterministic generative process in Lemma \ref{lemma:1} and establish its equivalence to the PF-ODE for the VE endpoint-fixed Schrödinger Bridge in Theorem \ref{theorem:1}. Proofs are provided in the \ref{appA2} and \ref{appA3}.
\begin{lemma}
If $g_n$ is set to 0, then equation (\ref{eq:i3sb_xn_2}) can be treated as an Euler discretization of the following ODE:
\begin{equation}
\text{d}\frac{X_t}{\sigma_t\overline{\sigma}_t}=\frac{X_1}{\sigma_1^2}\text{d}\frac{\sigma_t}{\overline{\sigma}_t}+\frac{\hat{X}_0^{\left(t\right)}\left(X_t\right)}{\sigma_1^2}\text{d}\frac{\overline{\sigma}_t}{\sigma_t}.
\label{eq:i3sb_ode}
\end{equation}
\label{lemma:1}
\end{lemma}
\begin{theorem}
The ODE (\ref{eq:i3sb_ode}) is equivalent to the PF-ODE for the VE endpoint fixed Schrödinger Bridge.
\label{theorem:1}
\end{theorem}

\subsection{Implementation Details}
We validated our proposed method using three groups of experiments: natural image, human face and medical image experiments.
\subsubsection{Natural Image Experiments}
For the natural image experiments, we used the pretrained I$^2$SB model and evaluated our proposed I$^3$SB on 10,000 images randomly selected from the validation set of ImageNet 256×256~\cite{deng2009imagenet}. We tested I$^3$SB on two image restoration tasks: 4× super-resolution with bicubic interpolation (sr4x-bicubic) and JPEG restoration with a quality factor of 10 (JPEG-10). The original images from the dataset served as clean images, and the corresponding corrupted images were generated by downsampling the clean images by a factor of 4 using bicubic interpolation for the sr4x-bicubic task, and by applying JPEG compression with a quality factor of 10 for the JPEG-10 task. The hyperparameter $\eta$ was set to 0.6.

\subsubsection{Human Face Experiments}
For the human face experiments, we conducted two groups of tests. In the first group, we used the CelebA-HQ~\cite{karras2017progressive} 512$\times$512 dataset, randomly splitting it into 27,000 images for training and 3,000 images for testing. In the second group, following SR3~\cite{saharia2022image}, we trained our model on the FFHQ~\cite{karras2019style} 512$\times$512 dataset and tested it on the CelebA-HQ 512$\times$512 dataset to assess its robustness to domain differences between the training and testing datasets. The entire FFHQ 512$\times$512 dataset was used for training, and 10,000 images were randomly selected from the CelebA-HQ 512$\times$512 dataset for testing. We evaluated our method on two image restoration tasks: sr4x-bicubic and JPEG-10. The original images from the dataset served as clean images, and the corresponding corrupted images were generated using the same procedure as in the natural image experiments. The hyperparameter $\eta$ was set to 0 in the sr4x-bicubic task in first group, and set to 0.2 in other tasks.

The neural network $\epsilon_{\theta}\left(X_n,t_n\right)$ we trained is a 2D residual U-Net with the same architecture used in DDPM~\cite{ho2020denoising}. We concatenated $X_{N}$ with $X_n$ along the channel dimension to serve as an additional condition for the network. During training, we used 1000 diffusion time steps with quadratic discretization, and adopted a symmetric scheduling of $\beta_t$~\cite{chen2022likelihood}. The model was trained on randomly cropped patches of size 128$\times$128 and tested on the entire 512$\times$512 images. A batch size of 64 was employed during training, using the Adam algorithm with a learning rate of $8\times10^{-5}$ for 200,000 iterations. 

\subsubsection{Medical Image Experiments}
For the medical image experiments, we evaluated our method on CT sparse view reconstruction, 4$\times$ super-resolution (sr4x) and denoising tasks. For CT sparse view reconstruction and sr4x tasks, we used the RPLHR-CT-tiny dataset~\cite{yu2022rplhr}, consisting of anonymized chest CT volumes. The original CT images served as clean images, and the corresponding corrupted images were generated using the FBP algorithm with projections from 60 distinct views in a fan beam geometry for the CT sparse view reconstruction task, and by downsampling the clean images by a factor of 4 in the projection domain for the CT sr4x task. We used 40 cases (11,090 slices) for training and 5 cases (1,425 slices) for testing. For the CT denoising task, we utilized the Mayo Grand Challenge dataset~\cite{Mayo_challenge}, which includes anonymized abdominal CT scans from 10 patients (5,936 slices) with matched full-dose (FD) data and simulated quarter-dose (QD) data. The FD data served as clean images, and the corresponding QD data served as corrupted images. We used data from 8 patients for training and 2 patients for testing in our experiment.  
We trained the neural network $\epsilon_{\theta}\left(X_n,t_n\right)$ using the same architecture and hyperparameters as in the human face experiments, with $\eta$ set to 0 during inference. 
\section{Results}
\subsection{Quantitative Results}
\subsubsection{Quantitative Results for Natural Image Experiments}
\begin{table}[t]
\begin{center}
\setlength{\tabcolsep}{2.4mm}{
\begin{tabular}{lccccc}
\hline
Method      &Time (s)&PSNR $\uparrow$       &SSIM $\uparrow$&LPIPS $\downarrow$&FID $\downarrow$\\
\hline
ADM~\cite{dhariwal2021diffusion}      & 4.91          & 27.18       &0.7767 & 0.2484          & 13.906            \\
DDNM~\cite{wang2022zero}       & 2.77        & \underline{27.54}          & \underline{0.7826}   & 0.2514          & 13.997        \\
DDRM~\cite{kawar2022denoising}    & 0.56          &27.17        & 0.7740&0.2730            & 19.700              \\
$\Pi$GDM~\cite{song2023pseudoinverse} & 9.84&25.48&0.7321&\textbf{0.2130}&4.382\\
DPS~\cite{chung2022diffusion} & 121.65&25.40&0.6940&0.3178&10.251\\
\hline
I$^2$SB~\cite{liu20232} (NFE=25)&0.67&26.11&0.7279&0.2671&6.633\\
I$^2$SB~\cite{liu20232} (NFE=50)&1.38&25.72&0.7130&0.2625&5.060\\
I$^2$SB~\cite{liu20232} (NFE=100)&2.79&25.44&0.7009&0.2598&4.128\\
\hline
I$^3$SB (ours, NFE=1)$^{\dagger}$&0.03&\textbf{29.07}&\textbf{0.8240}&\underline{0.2368}&13.299\\
I$^3$SB (ours, NFE=25)&0.67&25.91&0.7150&0.2568&4.115\\
I$^3$SB (ours, NFE=50)&1.38&25.63&0.7043&0.2586&\underline{3.766}\\
I$^3$SB (ours, NFE=100)&2.79&25.49&0.6992&0.2600&\textbf{3.648}\\
\hline
\end{tabular}}
\caption{
Quantitative results and computation time (per image) for the sr4x-bicubic task in natural image experiments. $^\dagger$I$^3$SB and I$^2$SB yield identical results when NFE=1. \textbf{Bold}: best, \underline{under}: second best.}
\label{tab1}
\end{center}
\end{table}

\begin{table}[t]
\begin{center}
\setlength{\tabcolsep}{2.4mm}{
\begin{tabular}{lccccc}
\hline
Method      &Time (s)&PSNR $\uparrow$       &SSIM $\uparrow$&LPIPS $\downarrow$&FID $\downarrow$\\
\hline
DDRM~\cite{kawar2022jpeg}    & 5.42          &\underline{28.50}       & \underline{0.8175}&0.2955            & 19.977             \\
$\Pi$GDM~\cite{song2023pseudoinverse} & 11.39&25.82&0.7349&0.2711&6.137\\
\hline
I$^2$SB~\cite{liu20232} (NFE=25)&0.79&27.09&0.7736&0.2529&5.970\\
I$^2$SB~\cite{liu20232} (NFE=50)&1.63&26.81&0.7623&0.2503&4.682\\
I$^2$SB~\cite{liu20232} (NFE=100)&3.29&26.62&0.7530&0.2488&3.871\\
\hline
I$^3$SB (ours ,NFE=1)$^{\dagger}$&0.03&\textbf{29.39}&\textbf{0.8400}&0.2677&16.977\\
I$^3$SB (ours, NFE=25)&0.79&26.86&0.7597&\textbf{0.2448}&3.764\\
I$^3$SB (ours, NFE=50)&1.63&26.67&0.7520&\underline{0.2470}&\underline{3.471}\\
I$^3$SB (ours, NFE=100)&3.29&26.58&0.7481&0.2485&\textbf{3.357}\\
\hline
\end{tabular}}
\end{center}
\caption{
Quantitative results and computation time (per image) for the JPEG-10 task in natural image experiments. $^\dagger$I$^3$SB and I$^2$SB yield identical results when NFE=1. \textbf{Bold}: best, \underline{under}: second best.}
\label{tab2}
\end{table}

\begin{table}[t]

\begin{center}
\setlength{\tabcolsep}{1.4mm}{
\begin{tabular}{lcccccccc}
\hline
 &&\multicolumn{2}{c}{Face (CelebA)}&\multicolumn{2}{c}{Face (FFHQ)}&\multicolumn{3}{c}{Medical Image} \\
\cmidrule(r){3-4}\cmidrule(r){5-6}\cmidrule(r){7-9}
Method&NFE&Sr4x&JPEG-10&Sr4x&JPEG-10&Sparse&Sr4x&Denoise\\
\hline
&25&8.318&12.999&8.810&15.678&43.958&45.211&44.080\\
cDDPM~\cite{ho2020denoising}&50&5.655&11.043&6.371&13.109&36.890&41.325&36.835\\
&100&4.263&9.432&4.790&10.376&29.333&36.264&29.584\\
\hline
&25&4.571&10.565&3.925&12.671&42.376&43.053&42.092\\
cDDIM~\cite{song2020denoising}&50&3.410&8.623&3.021&10.293&34.301&39.231&35.357\\
&100&\underline{3.096}&7.115&\underline{2.725}&\textbf{8.475}&26.203&34.191&28.335\\
\hline
&25&6.361&7.330&5.257&13.245&36.607&38.982&29.570\\
I$^2$SB~\cite{liu20232}&50&4.768&6.126&3.689&11.468&28.888&32.538&22.207\\
&100&3.870&5.587&3.111&10.172&\underline{22.448}&\underline{26.551}&\underline{15.986}\\
\hline
&25&3.996&5.694&2.842&10.382&33.240&37.682&28.200\\
I$^3$SB (ours)&50&3.237&\textbf{5.338}&\textbf{2.693}&9.129&24.476&30.475&20.387\\
&100&\textbf{3.009}&\underline{5.475}&2.918&\underline{8.568}&\textbf{17.673}&\textbf{23.772}&\textbf{14.028}\\
\hline
\end{tabular}}
\end{center}
\caption{
FIDs for all tasks in human face and medical image experiments. "Face (CelebA)" refers to experiments trained and tested on CelebA-HQ, and "Face (FFHQ)" refers to experiments trained on FFHQ and tested on CelebA-HQ. "Sparse" represents the CT sparse view reconstruction task. Lower FID indicates better performance. \textbf{Bold}: best, \underline{Underlined}: second best.}
\label{tab3}
\end{table}

In natural image experiments, we compared I$^3$SB with several state-of-the-art methods, which can be grouped into three categories: (1) paired data Schrödinger Bridge, including I$^2$SB~\cite{liu20232}, (2) conditional diffusion models, such as ADM~\cite{dhariwal2021diffusion}, and (3) diffusion-based task-agnostic models, including DDNM~\cite{wang2022zero}, DDRM~\cite{kawar2022denoising,kawar2022jpeg}, $\Pi$GDM~\cite{song2023pseudoinverse}, and DPS~\cite{chung2022diffusion}. For quantitative evaluation, we used Frechet Inception Distance (FID)~\cite{heusel2017gans} and Learned Perceptual Image Patch Similarity (LPIPS)~\cite{zhang2018perceptual} to assess perceptual quality and texture restoration, and Peak Signal-to-Noise Ratio (PSNR) and Structural Similarity Index Measure (SSIM) to evaluate fidelity to the ground truth. The experimental results for the sr4x-bicubic task are shown in Table \ref{tab1}, and for the JPEG-10 task in Table \ref{tab2}. Baseline values were computed using the official implementations of these methods with default hyperparameters. All experiments were conducted on a single A100 GPU, with computation times for each method included in the tables for reference.

In the sr4x-bicubic task (Table \ref{tab1}), I$^3$SB achieved the highest PSNR and SSIM with NFE=1 and the best FID with NFE=100. When comparing I$^3$SB to I$^2$SB at the same NFE, I$^3$SB consistently attained better FID, with a 38\% reduction at NFE=25, a 26\% reduction at NFE=50, and a 12\% reduction at NFE=100. Notably, compared to 100-step I$^2$SB, the 25-step I$^3$SB offered a fourfold acceleration in computation time while achieving slightly better performance across all metrics, including a 0.5 improvement in PSNR, a 0.015 increase in SSIM, a 0.003 reduction in LPIPS, and a 0.01 reduction in FID. When compared to ADM, DDNM, and DDRM, I$^3$SB with a single generative step outperformed these methods, showing a PSNR increase of 1.5 to 2, a 0.04 to 0.05 improvement in SSIM, a 0.01 to 0.04 decrease in LPIPS, and an FID reduction of 0.6 to 6. Against DPS, the 25-step I$^3$SB demonstrated superior performance, with a 0.5 increase in PSNR, a 0.02 improvement in SSIM, a 19\% reduction in LPIPS, and a 60\% reduction in FID. While $\Pi$GDM achieved the best LPIPS in this task, it required a known forward operator to incorporate data consistency during inference, had a worse FID, and was 14 times slower than the 25-step I$^3$SB.

In the JPEG-10 task (Table \ref{tab2}), I$^3$SB achieved the highest PSNR and SSIM with NFE=1, the best LPIPS with NFE=25, and the best FID with NFE=100. Similar to the sr4x-bicubic task, I$^3$SB consistently outperformed I$^2$SB in FID at the same NFE, with a 37\% reduction at NFE=25, a 26\% reduction at NFE=50, and a 13\% reduction at NFE=100. Notably, compared to the 100-step I$^2$SB, the 25-step I$^3$SB provided a fourfold acceleration in computation time while achieving slightly better performance across all metrics, including a 0.2 increase in PSNR, a 0.007 improvement in SSIM, a 0.004 reduction in LPIPS, and a 0.1 reduction in FID.
When compared to DDRM, I$^3$SB with a single generative step delivered superior results, with a 0.9 increase in PSNR, a 0.02 improvement in SSIM, a 0.03 reduction in LPIPS, and a 3-point decrease in FID. Against $\Pi$GDM, the 25-step I$^3$SB demonstrated better performance, achieving a 1-point improvement in PSNR, a 0.025 increase in SSIM, a 10\% reduction in LPIPS, and a 40\% reduction in FID.
\subsubsection{Quantitative Results for Human Face and Medical Image Experiments}
In the human face and medical image experiments, we compared I$^3$SB with two groups of state-of-the-art methods: (1) paired data Schrödinger Bridges, such as I$^2$SB~\cite{liu20232}, and (2) conditional diffusion models, including conditional DDPM (cDDPM)~\cite{ho2020denoising} and conditional DDIM (cDDIM)~\cite{song2020denoising}. We implemented cDDPM and cDDIM ourselves, with implementation details provided in ~\ref{appB}. For each task, we evaluated the perceptual quality of the tested methods using FID at NFE values of 25, 50, and 100. The results are presented in Table \ref{tab3}.

In the human face experiments (Table \ref{tab3}), which included tasks trained on CelebA-HQ and FFHQ datasets with low-resolution and JPEG compression corruptions, I$^3$SB consistently outperformed I$^2$SB and cDDPM in terms of FID under the same NFEs. Compared to I$^2$SB, I$^3$SB achieved FID reductions of 22\%-46\% at NFE=25, 13\%-32\% at NFE=50, and 2\%-22\% at NFE=100. Notably, the 25-step I$^3$SB achieved FIDs comparable to the 100-step I$^2$SB, demonstrating a 4$\times$ speedup. When compared to cDDPM, I$^3$SB reduced FID by 34\%-68\% at NFE=25, 30\%-58\% at NFE=50, and 17\%-42\% at NFE=100. While I$^3$SB and cDDIM achieved similar FIDs at NFE=100, I$^3$SB delivered superior performance at lower NFEs, with FID reductions of 13\%-46\% at NFE=25 and 5\%-38\% at NFE=50.

In the medical image experiments (Table \ref{tab3}), which included CT sparse-view reconstruction, sr4x, and denoising tasks, I$^3$SB consistently outperformed I$^2$SB, cDDPM, and cDDIM in terms of FID under the same NFEs. Compared to I$^2$SB, I$^3$SB achieved FID reductions of 3\%–9\% at NFE=25, 6\%–15\% at NFE=50, and 12\%–21\% at NFE=100. When compared to cDDPM and cDDIM, I$^3$SB delivered FID reductions of 12\%–36\% at NFE=25, 22\%–45\% at NFE=50, and 30\%–53\% at NFE=100.

\subsection{FID-NFE and FID-SSIM Curves}
\begin{figure}[!t]
\centering
\includegraphics[width=4in]{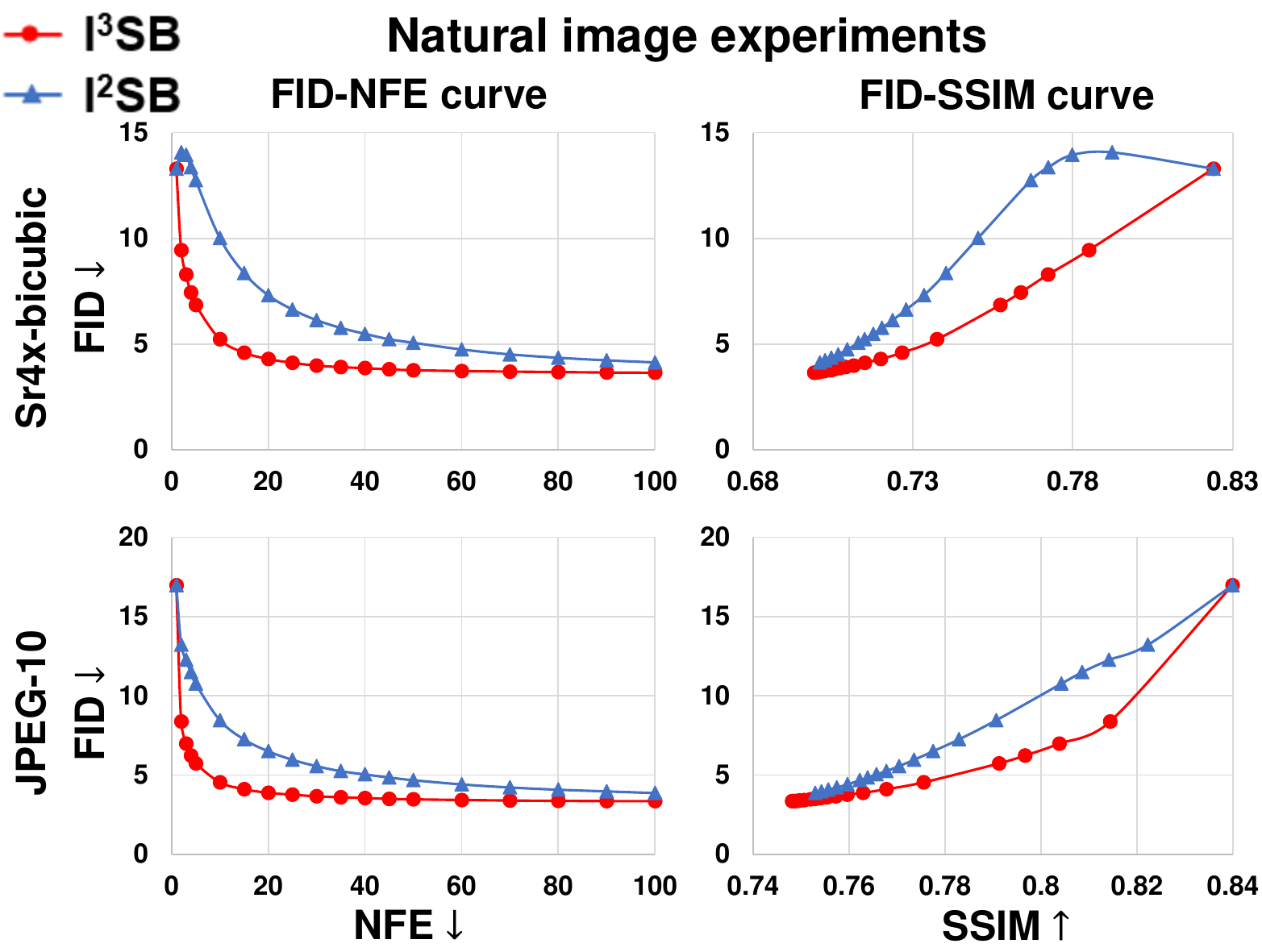}
\caption{FID-NFE and FID-SSIM curves for the sr4x-bicubic and JPEG-10 tasks in natural image experiments. Each point on the FID-SSIM curves represents the FID and SSIM values at a specific NFE, ranging from 1 to 100. As NFE increases, the FID-SSIM curves shift from the top-right to the bottom-left. Red curves correspond to I$^3$SB, and blue curves correspond to I$^2$SB.}
\label{fig:natural_curve}
\end{figure}
\begin{figure}[!t]
\centering
\includegraphics[width=4in]{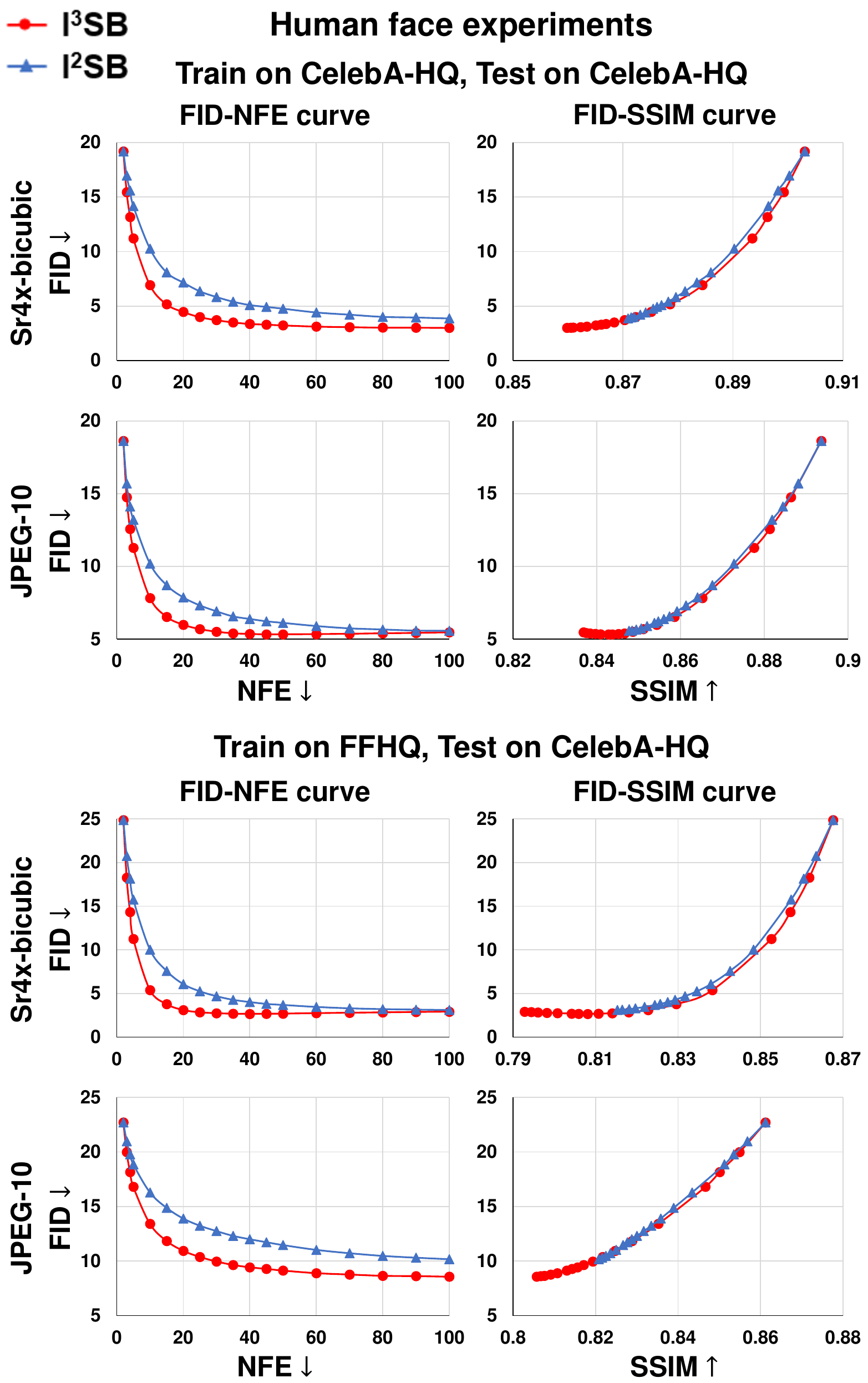}
\caption{FID-NFE and FID-SSIM curves for sr4x-bicubic and JPEG-10 tasks in human face experiments. Each point on the FID-SSIM curves represents the FID and SSIM values at a specific NFE, ranging from 2 to 100. As NFE increases, the FID-SSIM curves shift from the top-right to the bottom-left. Red curves correspond to I$^3$SB, and blue curves correspond to I$^2$SB.}
\label{fig:face_curve}
\end{figure}

\begin{figure}[!t]
\centering
\includegraphics[width=4in]{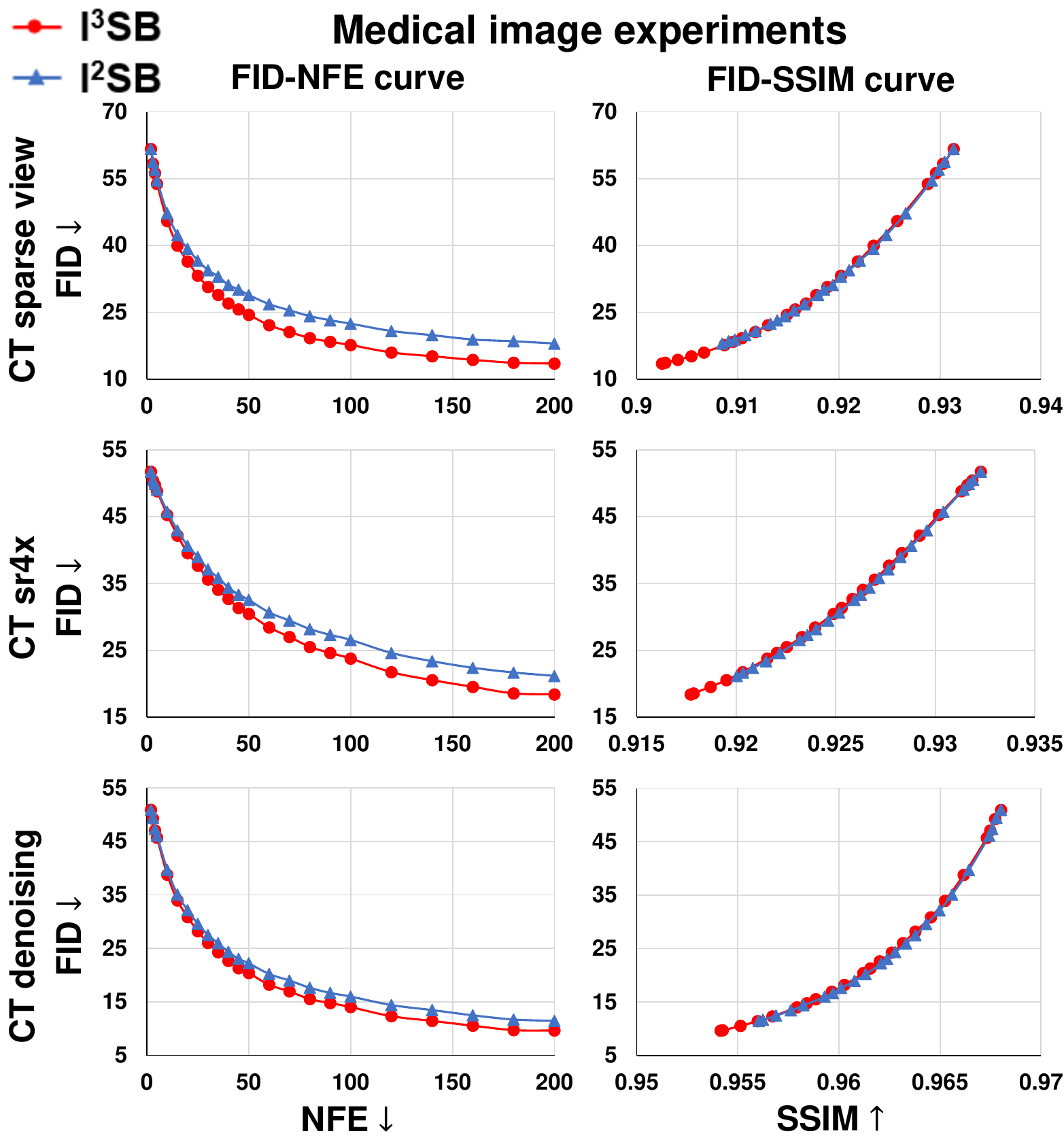}
\caption{FID-NFE and FID-SSIM curves for CT sparse view reconstruction, sr4x and denoising tasks in medical image experiments. Each point on the FID-SSIM curves represents the FID and SSIM values at a specific NFE, ranging from 2 to 200. As NFE increases, the FID-SSIM curves shift from the top-right to the bottom-left. Red curves correspond to I$^3$SB, and blue curves correspond to I$^2$SB.}
\label{fig:medical_curve}
\end{figure}
To further illustrate the acceleration effect of I$^3$SB, we plotted the FID-NFE and FID-SSIM curves for both I$^2$SB and I$^3$SB across all experiments. Each point on the FID-SSIM curves corresponds to the FID and SSIM values at a specific NFE. As NFE increases, the FID-SSIM curves shift from the top-right (high FID, high SSIM) to the bottom-left (low FID, low SSIM). These curves are presented in Figure \ref{fig:natural_curve} for natural image experiments, Figure \ref{fig:face_curve} for human face experiments, and Figure \ref{fig:medical_curve} for medical image experiments.

Consistent trends emerged across all the tasks for all the experiments. As shown in the FID-NFE curves, both I$^2$SB and I$^3$SB exhibited decreasing FID as NFE increased, indicating that higher NFEs improved perceptual quality for both models. Notably, FID decreased faster in I$^3$SB as NFE increased, meaning that I$^3$SB achieved the same perceptual quality with fewer generative steps.

The FID-SSIM curves highlighted the trade-off  between perceptual quality and fidelity to the ground truth, controlled by NFE. As NFE increases, the curves shift toward lower FID (better perceptual quality) but lower SSIM (reduced fidelity), reflecting the inherent compromise between detail generation and distortion~\cite{chung2024direct}. Notably, the FID-SSIM curves for I$^3$SB either overlapped with or shifted to the lower right of those for I$^2$SB, indicating that at equivalent perceptual quality, I$^3$SB maintained equal or better fidelity. This demonstrated that I$^3$SB accelerated the generative process of I$^2$SB, achieving the same perceptual quality with fewer generative steps, without introducing additional distortion from the ground truth.

Additionally, these trends were consistent when training on the FFHQ dataset and testing on the CelebA-HQ dataset in the human face experiments (Figure \ref{fig:face_curve}), indicating that the acceleration impact of I$^3$SB was robust to domain differences.

In natural image experiments, I$^3$SB achieved comparable performance to the 100-step I$^2$SB with 25 generative steps, resulting in a 4$\times$ speedup. For human face experiments, I$^3$SB matched the results of the 100-step I$^2$SB with 20 to 30 steps, providing a 3$\times$ to 5$\times$ speedup. In medical image experiments, I$^3$SB showed a 1.4$\times$ to 2$\times$ acceleration, achieving similar performance as the 200-step I$^2$SB with 100 steps in the CT sparse view reconstruction task, and 140 steps in both the CT sr4x and denoising tasks.
\subsection{Visualization Results}
\begin{figure}[!t]
\centering
\includegraphics[width=\textwidth]{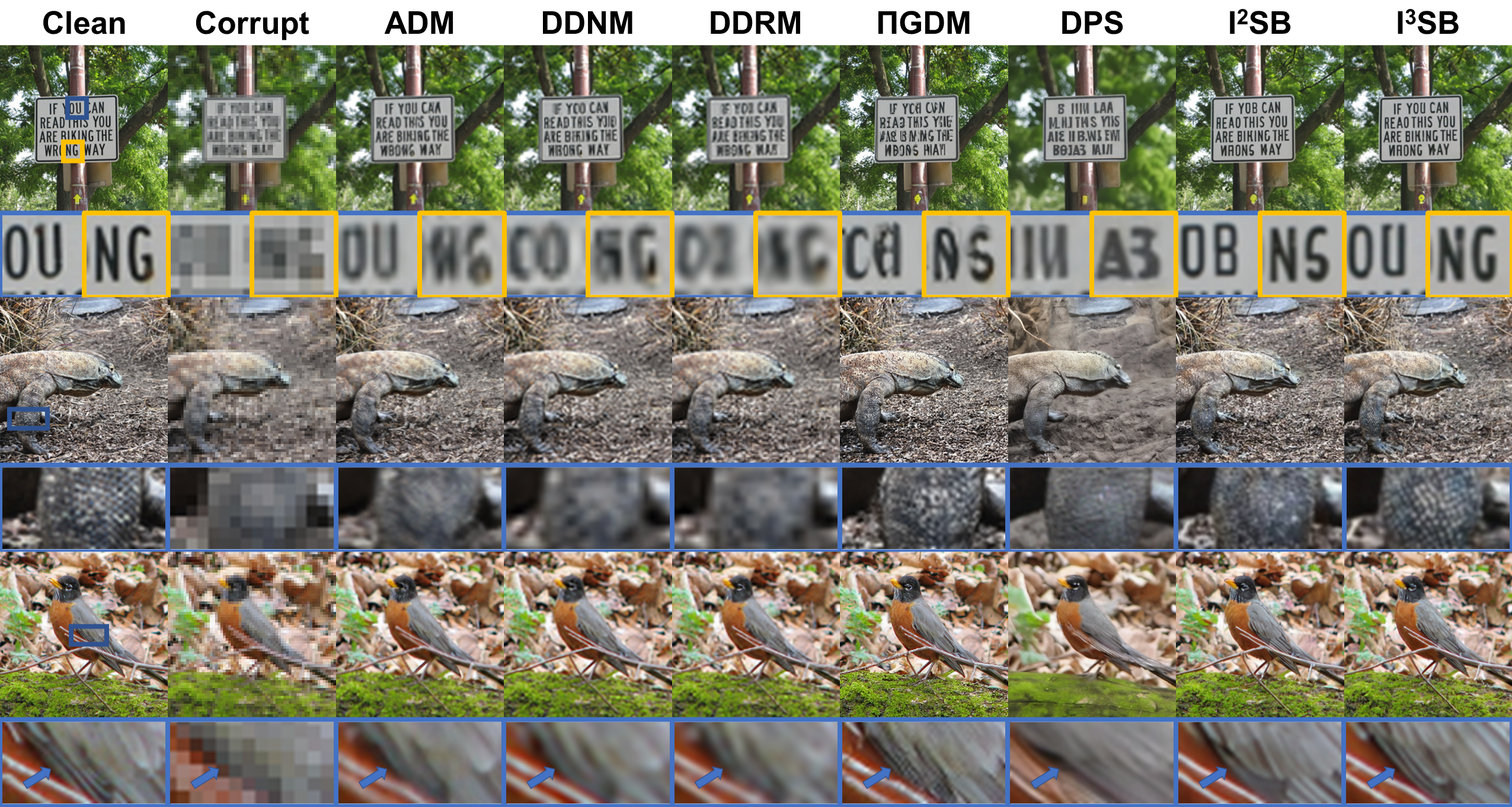}
\caption{Visualization result for the sr4x-bicubic task in the natural image experiments. Details within blue and yellow boxes are zoomed in for enhanced visual clarity. The NFE for I$^2$SB is 100, and for I$^3$SB is 25.}
\label{fig:natural_sr4x_img}
\end{figure}
\begin{figure}[!t]
\centering
\includegraphics[width=\textwidth]{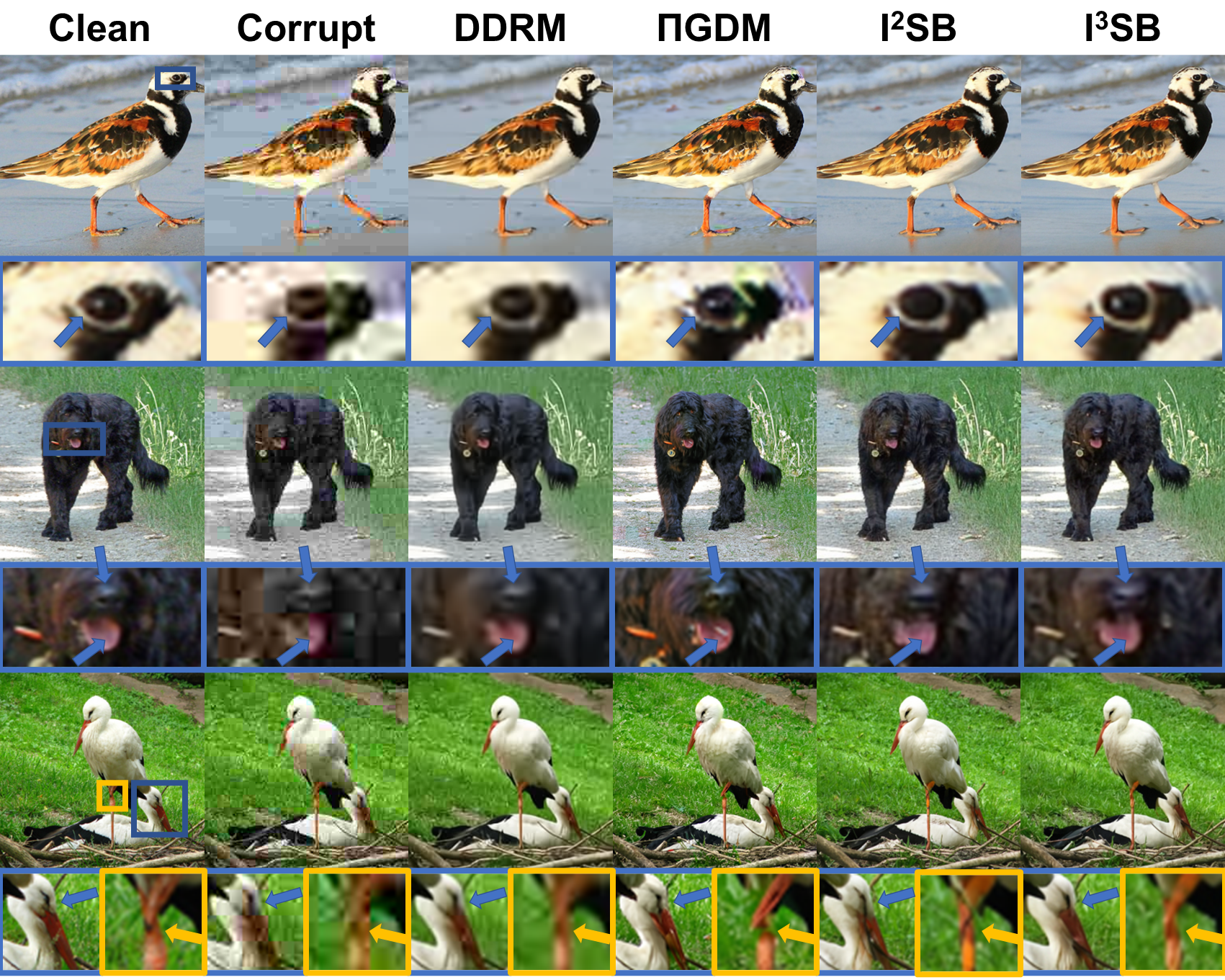}
\caption{Visualization results for the JPEG-10 task in the natural image experiments. Details within blue and yellow boxes are zoomed in for enhanced visual clarity. The NFE for I$^2$SB is 100, and for I$^3$SB is 25.}
\label{fig:natural_jpeg_img}
\end{figure}
\begin{figure}[!t]
\centering
\includegraphics[width=\textwidth]{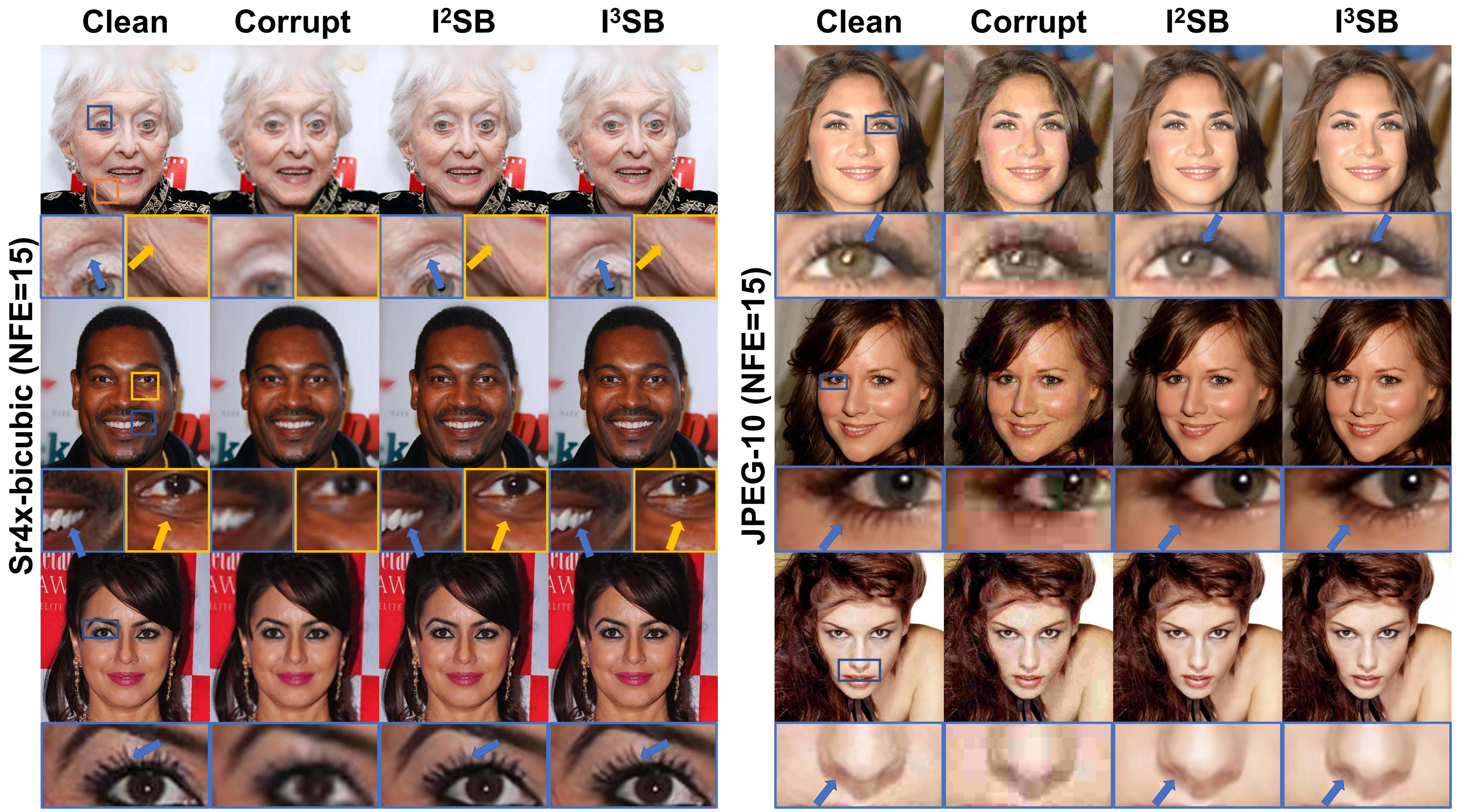}
\caption{Visualization results for sr4x-bicubic and JPEG-10 tasks in the human face experiments, trained and tested on CelebA-HQ. Details within blue and yellow boxes are zoomed in for enhanced visual clarity.}
\label{fig:face_image}
\end{figure}
\begin{figure}[!t]
\centering
\includegraphics[width=\textwidth]{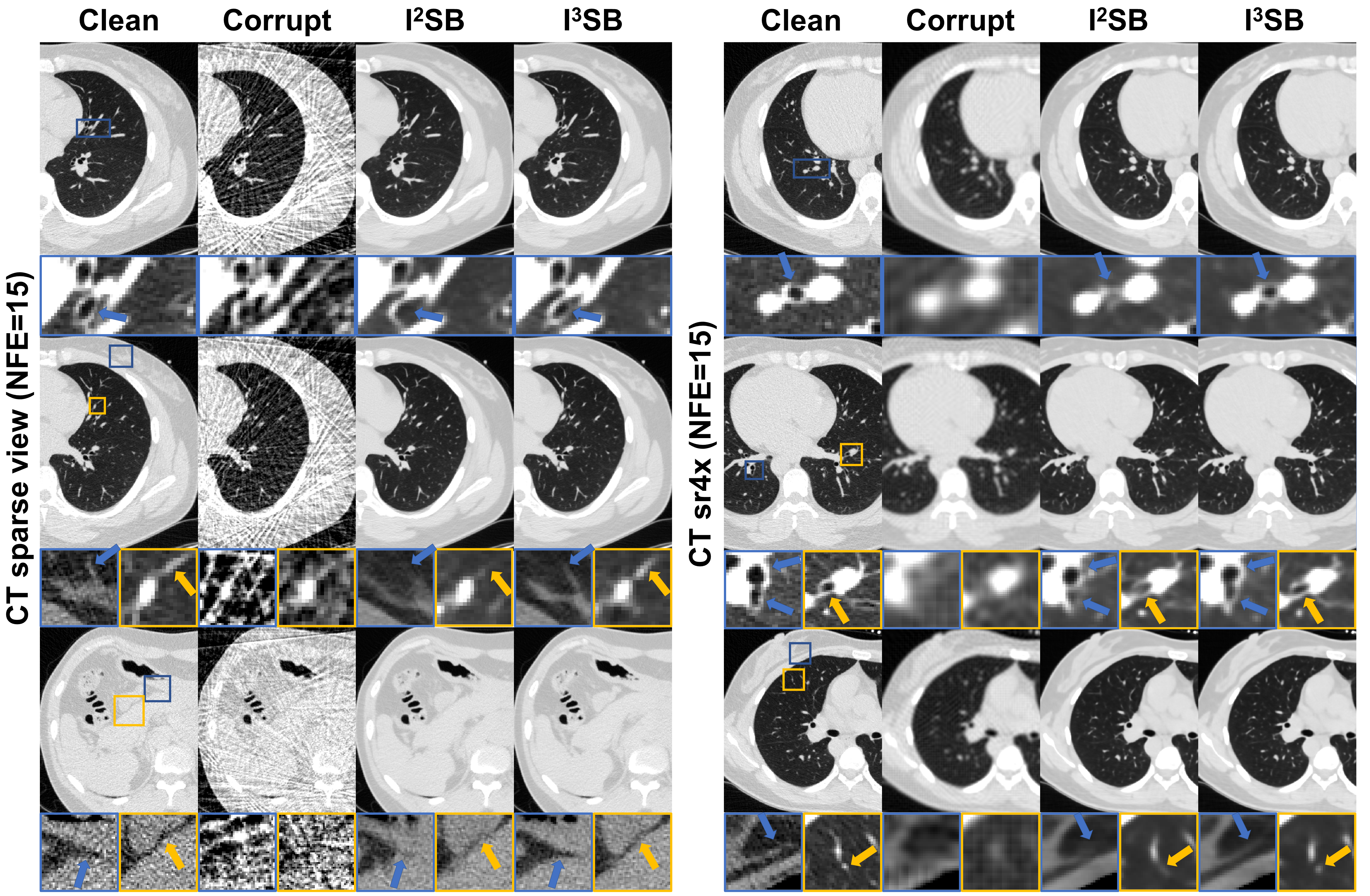}
\caption{Visualization results for the CT sparse view reconstruction and sr4x tasks in the medical image  experiments. The details within the blue and yellow boxes are zoomed in for enhanced visual clarity. The display window for the entire images is set to [-1000HU, 200HU], for the zoomed regions outside the lungs is set to [-160HU, 200HU], and for the zoomed regions inside the lungs is set to [-1000HU, -250HU].}
\label{fig:medical_image}
\end{figure}
The superior performance of I$^3$SB is further demonstrated by the visualization results in Figure \ref{fig:natural_sr4x_img} and Figure \ref{fig:natural_jpeg_img} for natural image experiments, Figure \ref{fig:face_image} for human face experiments, and Figure \ref{fig:medical_image} for medical image experiments. Compared to all comparison methods, I$^3$SB exhibitted enhanced detail restoration across all figures. Specifically, it excelled in restoring finer details, such as text characters, lizard scales, and bird features in Figure \ref{fig:natural_sr4x_img}; the bird's eyes and legs, as well as the dog's nose and tongue in Figure \ref{fig:natural_jpeg_img}, wrinkles, eyelashes, teeth, and nose in Figure \ref{fig:face_image}; and pulmonary veins and mammary glands in Figure \ref{fig:medical_image}.

\section{Conclusion and Discussion}
In conclusion, we introduce I$^3$SB to accelerate the generative process of I$^2$SB for image restoration tasks. By incorporating corrupted images into each generative step, I$^3$SB fully leverages their information, reformulating the process into a non-Markovian framework while maintaining the same marginal distribution as I$^2$SB, enabling direct use of pretrained models. A single hyperparameter balances the Markovian and non-Markovian components. Additionally, we establish the equivalence between I$^3$SB’s deterministic generative process and the PF-ODE of the VE endpoint-fixed Schrödinger Bridge. Extensive experiments across many image restoration tasks—including super-resolution and JPEG restoration for natural and human face images, as well as sparse-view reconstruction, super-resolution, and denoising for medical images—demonstrate the significant acceleration benefits of I$^3$SB. Compared to I$^2$SB, I$^3$SB achieves the same perceptual quality with fewer generative steps while maintaining or improving fidelity to the ground truth. Additionally, I$^3$SB outperforms several state-of-the-art diffusion-based image restoration models in both quantitative metrics and visual quality.

Despite these encouraging results, there are some limitations to our approach. First, I$^3$SB accelerates the generative process of I²SB by using the information present in the corrupted image. Consequently, in cases involving corruptions like occlusion, or extremely severe corruptions (e.g., 64× super-resolution), where the corrupted image contains little to no useful information, I$^3$SB may not achieve substantial acceleration. Second, similar to I$^2$SB, I$^3$SB requires paired clean and corrupted images for training, which limits its applicability in unpaired or weakly supervised settings. Third, I³SB implicitly assumes that clean images are noise-free. This assumption may hinder its performance when the clean images contain inherent noise, as is often the case in medical imaging. This is a potential reason why I$^3$SB achieves approximately 4× speedup in natural image and human face experiments but is limited to 2× speedup in medical image experiments. 

In future work, alongside addressing the limitations mentioned above, we plan to extend I$^3$SB in several directions. First, while I$^3$SB is currently based on the I$^2$SB framework, which operates as a Variance Exploding (VE) paired data Schrödinger Bridge, we aim to expand I$^3$SB to the Variance Preserving (VP) paired data Schrödinger Bridges. Second, since I$^3$SB offers an efficient deterministic sampling approach when $\eta=0$, it could serve as a foundation for distillation in paired data Schrödinger Bridges, potentially further accelerating the generative process. Third, as emphasized by CDDB~\cite{chung2024direct}, incorporating data consistency into the generative process—when the forward operator of the corruption is known—can enhance the performance of paired data Schrödinger Bridges. We plan to investigate this by integrating data consistency constraints into I$^3$SB, particularly for tasks like CT sparse view reconstruction, where the forward operator is available. We hope that I$^3$SB will serve as a robust and versatile framework for future research in image restoration, inspiring new advancements in the field.

\appendix
\section{Proofs}
\label{app1}
\subsection{Proof for Equations (\ref{eq:ABC}) and (\ref{eq:i3sb_xn_2})}
\label{appA1}
We define $p\left(X_{n-1}|X_0,X_N\right)$ as:
\begin{equation}
p\left(X_{n-1}|X_0,X_N\right)=\int p_G\left(X_{n-1}|X_0,X_{n},X_N\right)q\left(X_{n}|X_0,X_N\right)\text{d}X_{n}.
\end{equation}
According to Bishop (2006)~\cite{bishop2006pattern}, $p\left(X_{n-1}|X_0,X_N\right)$ is Gaussian, denoted as $\mathcal{N}\left(X_{n-1}|\mu_{n-1},\Sigma_{n-1}\right)$ where
\begin{subequations}
\begin{align}
\mu_{n-1}&=A_nX_0+C_nX_N+B_n\left(\frac{\overline{\sigma}_{n}^2}{\sigma_N^2}X_0+\frac{\sigma_{n}^2}{\sigma_N^2}X_N\right), \\
\Sigma_{n-1}&=(g_n^2+B_n^2\frac{\sigma_{n}^2\overline{\sigma}_{n}^2}{\sigma_N^2})I.
\end{align}
\end{subequations}
Given equation (\ref{eq:i3sb_sample}), $p\left(X_{n-1}|X_0,X_N\right)$ equals to $q\left(X_{n-1}|X_0,X_N\right)$
for any $X_{n-1}$, $X_0$ and $X_N$. Therefore, the weights for $X_0$ and $X_N$, as well as the covariance matrix of $p\left(X_{n-1}|X_0,X_N\right)$, should match those of $q\left(X_{n-1}|X_0,X_N\right)$. This leads to the following equations:
\begin{subequations}\label{eq:ABC_match}
\begin{align}
\frac{\overline{\sigma}_{n-1}^2}{\sigma_N^2}&=A_n+B_n\frac{\overline{\sigma}_{n}^2}{\sigma_N^2},\\
\frac{\sigma_{n-1}^2}{\sigma_N^2}&=C_n+B_n\frac{\sigma_{n}^2}{\sigma_N^2},\\
\frac{\overline{\sigma}_{n-1}^2\sigma_{n-1}^2}{\sigma_N^2}&=g_n^2+B_n^2\frac{\sigma_{n}^2\overline{\sigma}_{n}^2}{\sigma_N^2}.
\end{align}
\end{subequations}
By solving the system of equations in (\ref{eq:ABC_match}), we obtain the expressions for $A_n$, $B_n$ and $C_n$ as given in equation (\ref{eq:ABC}).

The sample $X_{n-1}$ is drawn from $p_G\left(X_{n-1}|\hat{X}_0^{(n)},X_{n},X_N\right)$ as in equation (\ref{eq:i3sb_xn}):
\begin{equation}
X_{n-1}=A_n\hat{X}_0^{(n)}+B_nX_{n}+C_nX_N+g_n\epsilon,
\label{eq:i3sb_xn3}
\end{equation}
where $\epsilon \sim \mathcal{N}\left(0,I\right)$ denotes Gaussian noise. By substituting the expressions for $A_n$, $B_n$ and $C_n$ into equation (\ref{eq:i3sb_xn3}), we derive the expression for $X_n$ as shown in equation (\ref{eq:i3sb_xn_2}).

\subsection{Proof for Lemma \ref{lemma:1}}
\label{appA2}
When $g_n$ is set to 0, equation (\ref{eq:i3sb_xn_2}) simplifies to:
\begin{equation}
X_{n-1}=\left(\frac{\overline{\sigma}_{n-1}^2}{\sigma_N^2}\hat{X}_0^{(n)}+\frac{\sigma_{n-1}^2}{\sigma_N^2}X_N\right)+\\\frac{\sigma_{n-1}\overline{\sigma}_{n-1}}{\sigma_{n}\overline{\sigma}_{n}}\left(X_{n}-\left(\frac{\overline{\sigma}_{n}^2}{\sigma_N^2}\hat{X}_0^{(n)}+\frac{\sigma_{n}^2}{\sigma_N^2}X_N\right)\right).
\end{equation}
This can be rearranged as:
\begin{equation}
\frac{X_{n-1}}{\sigma_{n-1}\overline{\sigma}_{n-1}}-\frac{X_{n}}{\sigma_{n}\overline{\sigma}_{n}}=\left(\frac{\sigma_{n-1}}{\overline{\sigma}_{n-1}}-\frac{\sigma_{n}}{\overline{\sigma}_{n}}\right)\frac{X_N}{\sigma_N^2}+\left(\frac{\overline{\sigma}_{n-1}}{\sigma_{n-1}}-\frac{\overline{\sigma}_{n}}{\sigma_{n}}\right)\frac{\hat{X}_0^{(n)}}{\sigma_N^2},
\end{equation}
which can be seen as an Euler discretization of the ODE in equation (\ref{eq:i3sb_ode}).
\subsection{Proof for Theorem \ref{theorem:1}}
\label{appA3}
To prove Theorem \ref{theorem:1}, we begin by deriving the PF-ODE for the VE endpoint-fixed Schrödinger Bridge. Following this, we establish its equivalence with the ODE given in equation (\ref{eq:i3sb_ode}).

The Schrödinger Bridge constructs diffusion bridges between two arbitrary distributions, $p_A$ and $p_B$. with the following forward and backward SDEs:
\begin{subequations}\label{eq:sde}
\begin{align}
\text{d}X_t&=\left[f_t+\beta_t\nabla_{X_t}\log\Psi\left(X_t,t\right)\right]\text{d}t+\sqrt{\beta_t}\text{d}w_t,\\
\text{d}X_t&=\left[f_t-\beta_t\nabla_{X_t}\log\hat{\Psi}\left(X_t,t\right)\right]\text{d}t+\sqrt{\beta_t}\text{d}\overline{w}_t.
\end{align}
\end{subequations}
Here, $X_0$ is sampled from $p_A$, $X_1$ is sampled from $p_B$, and $w_t$ and $\overline{w}_t$ denote the Wiener process and its time-reversed counterpart. To ensure that the path measure induced by the forward SDE (\ref{eq:sde}a) is almost surely equal to the one induced by the reverse SDE (\ref{eq:sde}b), the time-varying energy potentials $\Psi$ and $\hat{\Psi}$ should satisfy the following coupled partial differential equations (PDEs):
\begin{subequations}\label{eq:pde}
\begin{align}
\frac{\partial\Psi}{\partial t}&=-\nabla\Psi^\mathsf{T}f-\frac{1}{2}\beta\triangle\Psi,\\
\frac{\partial\hat{\Psi}}{\partial t}&=-\nabla\cdot\left(\hat{\Psi} f\right)+\frac{1}{2}\beta\triangle\hat{\Psi},
\end{align}
\end{subequations}
with the marginal conditions:
\begin{subequations}\label{eq:PDE_margin}
\begin{align}
\Psi\left(X_0,0\right)\hat{\Psi}\left(X_0,0\right)&=p_A\left(X_0\right),\\
\Psi\left(X_1,1\right)\hat{\Psi}\left(X_1,1\right)&=p_B\left(X_1\right).
\end{align}
\end{subequations}
As noted by Chen 2021~\cite{chen2022likelihood}, the ODE
\begin{equation}
\label{eq:probability_flow}
\text{d}X_t=\left(f_t+\frac{1}{2}\beta_t\nabla_{X_t}\log\frac{\Psi\left(X_t,t\right)}{\hat{\Psi}\left(X_t,t\right)}\right)\text{d}t
\end{equation}
characterizes the probability flow of the forward and reverse processes of the Schrödinger Bridge defined in equation (\ref{eq:sde}).

VE endpoint-fixed Schrödinger Bridge defines $f_t$, $p_A$ and $p_B$ as follows:
\begin{subequations}\label{eq:fpApB}
\begin{align}
f_t&=0,\\
p_A\left(X_0\right)&=q_{\text{clean}}\left(X_0|X_{\text{corrupt}}\right),\\
p_B\left(X_1\right)&=\delta\left(X_1-X_{\text{corrupt}}\right),
\end{align}
\end{subequations}
where $X_{\text{corrupt}}$ represents a given corrupted image, and $q_{\text{clean}}\left(\cdot|X_{\text{corrupt}}\right)$ represents the clean image distribution conditioned on $X_{\text{corrupt}}$. The Dirac function $\delta$ indicates that the endpoint fixed Schrödinger Bridge is constructed for each specific corrupted image, rather than for the entire corrupted image distribution.

\begin{lemma}
If $p_A$, $p_B$ and $f_t$ are defined as in equations (\ref{eq:fpApB}), then the PDEs (\ref{eq:pde}) with the marginal conditions (\ref{eq:PDE_margin}) admit the following analytical solutions:
\begin{equation}
\Psi\left(X_t,t\right)=\mathcal{N}\left(X_t|X_{\text{corrupt}},\overline{\sigma}_t^2I\right),
\label{eq:psi}
\end{equation}
\begin{equation}
\hat{\Psi}\left(X_t,t\right)=\int\hat{\Psi}_{X_0}\left(X_t,t\right)q_{\text{clean}}\left(X_0|X_{\text{corrupt}}\right)\text{d}X_0,
\label{eq:psi_hat}
\end{equation}
where
\begin{equation}
\hat{\Psi}_{X_0}\left(X_t,t\right)=C_{X_0}\mathcal{N}\left(X_t|X_0,\sigma_t^2I\right),
\label{eq:psi_hat_x0}
\end{equation}
and
\begin{equation}
C_{X_0}=\left(\sqrt{2\pi}\sigma_{1}\right)^d\exp{\left(\frac{\left(X_0-X_{\text{corrupt}}\right)^\mathsf{T}\left(X_0-X_{\text{corrupt}}\right)}{2\sigma_1^2}\right)}.
\label{eq:C_x0}
\end{equation}
Furthermore,
\begin{equation}
\nabla_{X_t}\log\Psi=-\frac{1}{\overline{\sigma}_t^2}\left(X_t-X_{\text{corrupt}}\right),
\label{eq:gradient_logpsi}
\end{equation}
and
\begin{equation}
\nabla_{X_t}\log\hat\Psi=-\frac{1}{\sigma_t^2}\left(X_t-\hat{X}_0^{(t)}\right),
\label{eq:gradient_logpsihat}
\end{equation}
where the expected mean $\hat{X}_0^{(t)}$ is defined as:
\begin{equation}
\hat{X}_0^{(t)}=\int X_0q_{\text{clean}}\left(X_0|X_t,X_{\text{corrupt}}\right)\text{d}X_0.
\end{equation}
\label{lemma:2}    
\end{lemma}

With Lemma \ref{lemma:2}, we provide proofs for the equivalence between the ODEs (\ref{eq:i3sb_ode}) and (\ref{eq:probability_flow}). Applying the chain rule, we obtain:
\begin{subequations}\label{eq:dsigma_sigmabar}
\begin{align}
\text{d}\frac{\sigma_t}{\overline{\sigma}_t}&=\frac{\sigma_1^2}{2\sigma_t\overline{\sigma}_t^3}\beta_t\text{d}t,\\
\text{d}\frac{\overline{\sigma}_t}{\sigma_t}&=-\frac{\sigma_1^2}{2\sigma_t^3\overline{\sigma}_t}\beta_t\text{d}t,
\end{align}
\end{subequations}
and
\begin{equation}\label{eq:dXt_sigma_sigmabar}
\begin{aligned}
\text{d}\frac{X_t}{\sigma_t\overline{\sigma}_t}&=\frac{1}{\sigma_t\overline{\sigma}_t}\text{d}X_t+X_t\text{d}\frac{1}{\sigma_t\overline{\sigma}_t},\\
&=\frac{1}{\sigma_t\overline{\sigma}_t}\text{d}X_t+X_t\frac{\sigma_t^2-\overline{\sigma}_t^2}{2\sigma_t^3\overline{\sigma}_t^3}\beta_t\text{d}t.
\end{aligned}
\end{equation}
Substituting equations (\ref{eq:dsigma_sigmabar}) and (\ref{eq:dXt_sigma_sigmabar}) into equation (\ref{eq:i3sb_ode}), we find that the ODE (\ref{eq:i3sb_ode}) is equivalent to:
\begin{equation}
\text{d}X_t=\frac{1}{2}\beta_t\left(\frac{1}{\overline{\sigma}_t^2}\left(X_1-X_t\right)+\frac{1}{\sigma_t^2}\left(X_t-\hat{X}_0^{(t)}\right)\right)\text{d}t.
\end{equation}
Using $X_1=X_{\text{corrupt}}$ along with equations (\ref{eq:fpApB}a), (\ref{eq:gradient_logpsi}) and (\ref{eq:gradient_logpsihat}), we conclude that the ODE (\ref{eq:i3sb_ode}) is equivalent to ODE (\ref{eq:probability_flow}), which is the PF- ODE for the VE endpoint fixed Schrödinger Bridge.
\subsection{Proof for Lemma \ref{lemma:2}}
To prove Lemma \ref{lemma:2}, we first demonstrate that $\Psi$, as defined in equation (\ref{eq:psi}), satisfies the PDE (\ref{eq:pde}a). This holds because:
\begin{equation}
\begin{aligned}
\frac{\partial\Psi}{\partial t} &= \frac{\partial{\Psi}}{\partial\overline{\sigma}_t^2}\frac{\partial\overline{\sigma}_t^2}{\partial t},\\
&=-\frac{1}{2}\beta\Psi\left(\frac{\left(X_t-X_{\text{corrupt}}\right)^\mathsf{T}\left(X_t-X_{\text{corrupt}}\right)}{\overline{\sigma}_t^4}-\frac{d}{\sigma_t^2}\right),\\
&=-\frac{1}{2}\beta\triangle\Psi.
\end{aligned}
\end{equation}
Similarly, $\hat\Psi_{X_0}\left(X_t,t\right)$ satisfies:$\frac{\partial\hat{\Psi}_{X_0}}{\partial t}=\frac{1}{2}\beta\triangle\hat{\Psi}_{X_0}$.
Consequently, $\hat{\Psi}$ expressed in equation (\ref{eq:psi_hat}) satisfies:
\begin{equation}
\begin{aligned}
\frac{\partial\hat{\Psi}}{\partial t}&=\int\frac{\partial\hat{\Psi}_{X_0}}{\partial t}q_{\text{clean}}\left(X_0|X_{\text{corrupt}},y\right)dX_0, \\
&=\frac{1}{2}\beta\int\triangle\hat{\Psi}_{X_0}q_{\text{clean}}\left(X_0|X_{\text{corrupt}},y\right)dX_0,\\
&=\frac{1}{2}\beta\triangle\hat{\Psi}.
\end{aligned}
\end{equation}
Therefore, $\Psi$ and $\hat{\Psi}$ satisfy the PDEs (\ref{eq:pde}). 

Next, we proof that $\Psi$ and $\hat{\Psi}$ satisfy the equations (\ref{eq:PDE_margin}). At $t=0$, we have
\begin{equation}
\Psi\left(X_0,0\right)=1/C_{X_0},
\end{equation}
where $C_{X_0}$ is defined in equation (\ref{eq:C_x0}). Additionally,
\begin{equation}
\begin{aligned}
\hat\Psi\left(X_0,0\right)&=\int C_X\delta\left(X_0-X\right)q_{\text{clean}}\left(X|X_{\text{corrupt}}\right)\text{d}X, \\
&=C_{X_0}q_{\text{clean}}\left(X_0|X_{\text{corrupt}}\right).
\end{aligned}
\end{equation}
Therefore, the marginal condition (\ref{eq:PDE_margin}a) holds. At $t=1$, we have
\begin{equation}
\Psi\left(X_1,1\right)=\delta\left(X_1-X_{\text{corrupt}}\right),
\end{equation}
and since $\hat{\Psi}_{X_0}\left(X_1=X_{\text{corrupt}},1\right)$ equals 1, we have:
\begin{equation}
\begin{aligned}
&\Psi\left(X_1,1\right)\hat{\Psi}\left(X_1,1\right)\\&=\delta\left(X_1-X_{\text{corrupt}}\right)\int\hat{\Psi}_{X_0}\left(X_1,1\right)q_{\text{clean}}\left(X_0|X_{\text{corrupt}}\right)\text{d}X_0, \\
&=\delta\left(X_1-X_{\text{corrupt}}\right)\int q_{\text{clean}}\left(X_0|X_{\text{corrupt}}\right)\text{d}X_0, \\
&=\delta\left(X_1-X_{\text{corrupt}}\right), \\
&=p_B\left(X_1\right).
\end{aligned}
\end{equation}
Thus, $\Psi$ and $\hat{\Psi}$ satisfy the marginal conditions (\ref{eq:PDE_margin}).

Finally, we provide proofs for equations (\ref{eq:gradient_logpsi}) and (\ref{eq:gradient_logpsihat}). For equation (\ref{eq:gradient_logpsi}), $\nabla_{X_t}\log\Psi$ can be straightforwardly obtained by direct computation. For equation (\ref{eq:gradient_logpsihat}), we proceed as follows:
\begin{equation} \label{eq:gradient_logpsihat2}
\begin{aligned}
\nabla\log\hat\Psi&=\frac{\nabla\hat\Psi}{\hat\Psi}\\
&=\frac{1}{\hat{\Psi}}\int\nabla\hat{\Psi}_{X_0}\left(X_t,t\right)q_{\text{clean}}\left(X_0|X_{\text{corrupt}}\right)\text{d}X_0\\
&=\frac{1}{\hat{\Psi}}\int\left(\frac{X_0-X_t}{\sigma_t^2}\right)\hat{\Psi}_{X_0}q_{\text{clean}}\left(X_0|X_{\text{corrupt}}\right)\text{d}X_0\\
&=-\frac{1}{\sigma_t^2}\left(X_t-\frac{1}{\hat{\Psi}}\int X_0\hat{\Psi}_{X_0}q_{\text{clean}}\left(X_0|X_{\text{corrupt}}\right)\text{d}X_0\right)
\end{aligned}
\end{equation}
Using the definition of $\hat\Psi_{X_0}$ in equation (\ref{eq:psi_hat_x0}), we have:
\begin{equation}
\hat\Psi_{X_0}\left(X_t,t\right)=k_{X_t}q\left(X_t|X_0,X_1=X_{\text{corrupt}}\right),
\end{equation}
where $q\left(X_t|X_0,X_1\right)$ is defined in equation (\ref{eq:mu_t}), and $k_{X_t}$ is independent of $X_0$. Since $X_1$ is sampled from a Dirac distribution centered at $X_{\text{corrupt}}$, we obtain:
\begin{equation}
q\left(X_t|X_0,X_1=X_{\text{corrupt}}\right)=q\left(X_t|X_0,X_{\text{corrupt}}\right).
\end{equation}
Therefore:
\begin{equation}
\nabla\log\hat\Psi=-\frac{1}{\sigma_t^2}\left(X_t-\frac{\int X_0q\left(X_t|X_0,X_{\text{corrupt}}\right)q_{\text{clean}}\left(X_0|X_{\text{corrupt}}\right)\text{d}X_0}{\int q\left(X_t|X_0,X_{\text{corrupt}}\right)q_{\text{clean}}\left(X_0|X_{\text{corrupt}}\right)\text{d}X_0}\right)
\end{equation}
Using Bayes' theorem, we know that:
\begin{equation}
q\left(X_t|X_0,X_{\text{corrupt}}\right)=\frac{q_{\text{clean}}\left(X_0|X_t,X_{\text{corrupt}}\right)q\left(X_t|X_{\text{corrupt}}\right)}{q_{\text{clean}}\left(X_0|X_{\text{corrupt}}\right)}
.
\end{equation}
Thus, we have
\begin{equation}\label{eq:X0_hat2}
\begin{aligned}
\nabla\log\hat\Psi&=-\frac{1}{\sigma_t^2}\left(X_t-\frac{\int X_0q_{\text{clean}}\left(X_0|X_t,X_{\text{corrupt}}\right)\text{d}X_0}{\int q_{\text{clean}}\left(X_0|X_t,X_{\text{corrupt}}\right)\text{d}X_0}\right)\\
&=-\frac{1}{\sigma_t^2}\left(X_t-\hat{X}_0^{(t)}\right).
\end{aligned}
\end{equation}
That completes the proof.
\section{Implementation Details for cDDPM and cDDIM}
\label{appB}
We implemented cDDPM and cDDIM as comparison methods for human face and medical image experiments. We adopted the sigmoid schedule~\cite{jabri2023scalable} for $\beta_t$ and used the same network architecture and training settings as the I$^2$SB model trained for these tasks. The network was trained to predict the ground truth in human face experiments and to predict the noise in medical image experiments. The hyperparameter $\eta$ in cDDIM was set to 0 for human face experiments and 0.6 for medical image experiments.
\section*{Declaration of generative AI and AI-assisted technologies in the writing process}
During the preparation of this work the authors used ChatGPT in order to improve language and readability. After using this tool, the authors reviewed and edited the content as needed and take full responsibility for the content of the publication.
\section*{Acknowledgements}
The authors acknowledge financial support provided by National Institute of Biomedical Imaging and Bioengineering and Samsung Electronics Co., Ltd.

%\bibliographystyle{elsarticle-num-names} 
%\bibliography{i3sb}

\begin{thebibliography}{40}
\expandafter\ifx\csname natexlab\endcsname\relax\def\natexlab#1{#1}\fi
\providecommand{\url}[1]{\texttt{#1}}
\providecommand{\href}[2]{#2}
\providecommand{\path}[1]{#1}
\providecommand{\DOIprefix}{doi:}
\providecommand{\ArXivprefix}{arXiv:}
\providecommand{\URLprefix}{URL: }
\providecommand{\Pubmedprefix}{pmid:}
\providecommand{\doi}[1]{\href{http://dx.doi.org/#1}{\path{#1}}}
\providecommand{\Pubmed}[1]{\href{pmid:#1}{\path{#1}}}
\providecommand{\bibinfo}[2]{#2}
\ifx\xfnm\relax \def\xfnm[#1]{\unskip,\space#1}\fi
%Type = Article
\bibitem[{Dhariwal and Nichol(2021)}]{dhariwal2021diffusion}
\bibinfo{author}{P.~Dhariwal}, \bibinfo{author}{A.~Nichol},
\newblock \bibinfo{title}{Diffusion models beat gans on image synthesis},
\newblock \bibinfo{journal}{Advances in neural information processing systems} \bibinfo{volume}{34} (\bibinfo{year}{2021}) \bibinfo{pages}{8780--8794}.
%Type = Article
\bibitem[{Saharia et~al.(2022)Saharia, Ho, Chan, Salimans, Fleet, and Norouzi}]{saharia2022image}
\bibinfo{author}{C.~Saharia}, \bibinfo{author}{J.~Ho}, \bibinfo{author}{W.~Chan}, \bibinfo{author}{T.~Salimans}, \bibinfo{author}{D.~J. Fleet}, \bibinfo{author}{M.~Norouzi},
\newblock \bibinfo{title}{Image super-resolution via iterative refinement},
\newblock \bibinfo{journal}{IEEE Transactions on Pattern Analysis and Machine Intelligence} \bibinfo{volume}{45} (\bibinfo{year}{2022}) \bibinfo{pages}{4713--4726}.
%Type = Article
\bibitem[{Gonzalez-Sabbagh et~al.(2024)Gonzalez-Sabbagh, Robles-Kelly, and Gao}]{gonzalez2024dgd}
\bibinfo{author}{S.~Gonzalez-Sabbagh}, \bibinfo{author}{A.~Robles-Kelly}, \bibinfo{author}{S.~Gao},
\newblock \bibinfo{title}{Dgd-cgan: A dual generator for image dewatering and restoration},
\newblock \bibinfo{journal}{Pattern Recognition} \bibinfo{volume}{148} (\bibinfo{year}{2024}) \bibinfo{pages}{110159}.
%Type = Inproceedings
\bibitem[{Liu et~al.(2023)Liu, Vahdat, Huang, Theodorou, Nie, and Anandkumar}]{liu20232}
\bibinfo{author}{G.-H. Liu}, \bibinfo{author}{A.~Vahdat}, \bibinfo{author}{D.-A. Huang}, \bibinfo{author}{E.~A. Theodorou}, \bibinfo{author}{W.~Nie}, \bibinfo{author}{A.~Anandkumar},
\newblock \bibinfo{title}{I2sb: Image-to-image schr{\"o}dinger bridge},
\newblock in: \bibinfo{booktitle}{International Conference on Machine Learning}, \bibinfo{year}{2023}. \URLprefix \url{https://api.semanticscholar.org/CorpusID:257022338}.
%Type = Article
\bibitem[{Chung et~al.(2024)Chung, Kim, and Ye}]{chung2024direct}
\bibinfo{author}{H.~Chung}, \bibinfo{author}{J.~Kim}, \bibinfo{author}{J.~C. Ye},
\newblock \bibinfo{title}{Direct diffusion bridge using data consistency for inverse problems},
\newblock \bibinfo{journal}{Advances in Neural Information Processing Systems} \bibinfo{volume}{36} (\bibinfo{year}{2024}).
%Type = Article
\bibitem[{Delbracio and Milanfar(2023)}]{delbracio2023inversion}
\bibinfo{author}{M.~Delbracio}, \bibinfo{author}{P.~Milanfar},
\newblock \bibinfo{title}{Inversion by direct iteration: An alternative to denoising diffusion for image restoration},
\newblock \bibinfo{journal}{Transactions on Machine Learning Research}  (\bibinfo{year}{2023}). \URLprefix \url{https://openreview.net/forum?id=VmyFF5lL3F}, \bibinfo{note}{featured Certification}.
%Type = Article
\bibitem[{Ho et~al.(2020)Ho, Jain, and Abbeel}]{ho2020denoising}
\bibinfo{author}{J.~Ho}, \bibinfo{author}{A.~Jain}, \bibinfo{author}{P.~Abbeel},
\newblock \bibinfo{title}{Denoising diffusion probabilistic models},
\newblock \bibinfo{journal}{Advances in neural information processing systems} \bibinfo{volume}{33} (\bibinfo{year}{2020}) \bibinfo{pages}{6840--6851}.
%Type = Inproceedings
\bibitem[{Song et~al.(2021)Song, Meng, and Ermon}]{song2020denoising}
\bibinfo{author}{J.~Song}, \bibinfo{author}{C.~Meng}, \bibinfo{author}{S.~Ermon},
\newblock \bibinfo{title}{Denoising diffusion implicit models},
\newblock in: \bibinfo{booktitle}{International Conference on Learning Representations}, \bibinfo{year}{2021}. \URLprefix \url{https://openreview.net/forum?id=St1giarCHLP}.
%Type = Inproceedings
\bibitem[{Zhou et~al.(2024)Zhou, Lou, Khanna, and Ermon}]{zhou2023denoising}
\bibinfo{author}{L.~Zhou}, \bibinfo{author}{A.~Lou}, \bibinfo{author}{S.~Khanna}, \bibinfo{author}{S.~Ermon},
\newblock \bibinfo{title}{Denoising diffusion bridge models},
\newblock in: \bibinfo{booktitle}{The Twelfth International Conference on Learning Representations}, \bibinfo{year}{2024}. \URLprefix \url{https://openreview.net/forum?id=FKksTayvGo}.
%Type = Article
\bibitem[{Buades et~al.(2011)Buades, Coll, and Morel}]{buades2011non}
\bibinfo{author}{A.~Buades}, \bibinfo{author}{B.~Coll}, \bibinfo{author}{J.-M. Morel},
\newblock \bibinfo{title}{Non-local means denoising},
\newblock \bibinfo{journal}{Image Processing On Line} \bibinfo{volume}{1} (\bibinfo{year}{2011}) \bibinfo{pages}{208--212}.
%Type = Article
\bibitem[{El~Hamidi et~al.(2010)El~Hamidi, Menard, Lugiez, and Ghannam}]{el2010weighted}
\bibinfo{author}{A.~El~Hamidi}, \bibinfo{author}{M.~Menard}, \bibinfo{author}{M.~Lugiez}, \bibinfo{author}{C.~Ghannam},
\newblock \bibinfo{title}{Weighted and extended total variation for image restoration and decomposition},
\newblock \bibinfo{journal}{Pattern Recognition} \bibinfo{volume}{43} (\bibinfo{year}{2010}) \bibinfo{pages}{1564--1576}.
%Type = Article
\bibitem[{Li et~al.(2022)Li, Liang, Zhang, Fan, and Yu}]{li2022joint}
\bibinfo{author}{P.~Li}, \bibinfo{author}{J.~Liang}, \bibinfo{author}{M.~Zhang}, \bibinfo{author}{W.~Fan}, \bibinfo{author}{G.~Yu},
\newblock \bibinfo{title}{Joint image denoising with gradient direction and edge-preserving regularization},
\newblock \bibinfo{journal}{Pattern Recognition} \bibinfo{volume}{125} (\bibinfo{year}{2022}) \bibinfo{pages}{108506}.
%Type = Article
\bibitem[{Aharon et~al.(2006)Aharon, Elad, and Bruckstein}]{aharon2006k}
\bibinfo{author}{M.~Aharon}, \bibinfo{author}{M.~Elad}, \bibinfo{author}{A.~Bruckstein},
\newblock \bibinfo{title}{K-svd: An algorithm for designing overcomplete dictionaries for sparse representation},
\newblock \bibinfo{journal}{IEEE Transactions on signal processing} \bibinfo{volume}{54} (\bibinfo{year}{2006}) \bibinfo{pages}{4311--4322}.
%Type = Inproceedings
\bibitem[{Zamir et~al.(2022)Zamir, Arora, Khan, Hayat, Khan, and Yang}]{zamir2022restormer}
\bibinfo{author}{S.~W. Zamir}, \bibinfo{author}{A.~Arora}, \bibinfo{author}{S.~Khan}, \bibinfo{author}{M.~Hayat}, \bibinfo{author}{F.~S. Khan}, \bibinfo{author}{M.-H. Yang},
\newblock \bibinfo{title}{Restormer: Efficient transformer for high-resolution image restoration},
\newblock in: \bibinfo{booktitle}{Proceedings of the IEEE/CVF conference on computer vision and pattern recognition}, \bibinfo{year}{2022}, pp. \bibinfo{pages}{5728--5739}.
%Type = Inproceedings
\bibitem[{Harvey et~al.(2022)Harvey, Naderiparizi, and Wood}]{harvey2022conditional}
\bibinfo{author}{W.~Harvey}, \bibinfo{author}{S.~Naderiparizi}, \bibinfo{author}{F.~Wood},
\newblock \bibinfo{title}{Conditional image generation by conditioning variational auto-encoders},
\newblock in: \bibinfo{booktitle}{International Conference on Learning Representations}, \bibinfo{year}{2022}. \URLprefix \url{https://openreview.net/forum?id=7MV6uLzOChW}.
%Type = Inproceedings
\bibitem[{Helminger et~al.(2021)Helminger, Bernasconi, Djelouah, Gross, and Schroers}]{helminger2021generic}
\bibinfo{author}{L.~Helminger}, \bibinfo{author}{M.~Bernasconi}, \bibinfo{author}{A.~Djelouah}, \bibinfo{author}{M.~Gross}, \bibinfo{author}{C.~Schroers},
\newblock \bibinfo{title}{Generic image restoration with flow based priors},
\newblock in: \bibinfo{booktitle}{Proceedings of the IEEE/CVF Conference on Computer Vision and Pattern Recognition}, \bibinfo{year}{2021}, pp. \bibinfo{pages}{334--343}.
%Type = Article
\bibitem[{Liu et~al.(2024)Liu, He, Liu, Lin, Yu, Hu, Qiao, and Dong}]{liu2024adaptbir}
\bibinfo{author}{Y.~Liu}, \bibinfo{author}{J.~He}, \bibinfo{author}{Y.~Liu}, \bibinfo{author}{X.~Lin}, \bibinfo{author}{F.~Yu}, \bibinfo{author}{J.~Hu}, \bibinfo{author}{Y.~Qiao}, \bibinfo{author}{C.~Dong},
\newblock \bibinfo{title}{Adaptbir: Adaptive blind image restoration with latent diffusion prior for higher fidelity},
\newblock \bibinfo{journal}{Pattern Recognition}  (\bibinfo{year}{2024}) \bibinfo{pages}{110659}.
%Type = Inproceedings
\bibitem[{Song et~al.(2021)Song, Sohl-Dickstein, Kingma, Kumar, Ermon, and Poole}]{song2020score}
\bibinfo{author}{Y.~Song}, \bibinfo{author}{J.~Sohl-Dickstein}, \bibinfo{author}{D.~P. Kingma}, \bibinfo{author}{A.~Kumar}, \bibinfo{author}{S.~Ermon}, \bibinfo{author}{B.~Poole},
\newblock \bibinfo{title}{Score-based generative modeling through stochastic differential equations},
\newblock in: \bibinfo{booktitle}{International Conference on Learning Representations}, \bibinfo{year}{2021}. \URLprefix \url{https://openreview.net/forum?id=PxTIG12RRHS}.
%Type = Inproceedings
\bibitem[{Wang et~al.(2023)Wang, Yu, and Zhang}]{wang2022zero}
\bibinfo{author}{Y.~Wang}, \bibinfo{author}{J.~Yu}, \bibinfo{author}{J.~Zhang},
\newblock \bibinfo{title}{Zero-shot image restoration using denoising diffusion null-space model},
\newblock in: \bibinfo{booktitle}{The Eleventh International Conference on Learning Representations}, \bibinfo{year}{2023}. \URLprefix \url{https://openreview.net/forum?id=mRieQgMtNTQ}.
%Type = Article
\bibitem[{Kawar et~al.(2022{\natexlab{a}})Kawar, Elad, Ermon, and Song}]{kawar2022denoising}
\bibinfo{author}{B.~Kawar}, \bibinfo{author}{M.~Elad}, \bibinfo{author}{S.~Ermon}, \bibinfo{author}{J.~Song},
\newblock \bibinfo{title}{Denoising diffusion restoration models},
\newblock \bibinfo{journal}{Advances in Neural Information Processing Systems} \bibinfo{volume}{35} (\bibinfo{year}{2022}{\natexlab{a}}) \bibinfo{pages}{23593--23606}.
%Type = Inproceedings
\bibitem[{Kawar et~al.(2022{\natexlab{b}})Kawar, Song, Ermon, and Elad}]{kawar2022jpeg}
\bibinfo{author}{B.~Kawar}, \bibinfo{author}{J.~Song}, \bibinfo{author}{S.~Ermon}, \bibinfo{author}{M.~Elad},
\newblock \bibinfo{title}{Jpeg artifact correction using denoising diffusion restoration models},
\newblock in: \bibinfo{booktitle}{Neural Information Processing Systems (NeurIPS) Workshop on Score-Based Methods}, \bibinfo{year}{2022}{\natexlab{b}}.
%Type = Inproceedings
\bibitem[{Chung et~al.(2023)Chung, Kim, Mccann, Klasky, and Ye}]{chung2022diffusion}
\bibinfo{author}{H.~Chung}, \bibinfo{author}{J.~Kim}, \bibinfo{author}{M.~T. Mccann}, \bibinfo{author}{M.~L. Klasky}, \bibinfo{author}{J.~C. Ye},
\newblock \bibinfo{title}{Diffusion posterior sampling for general noisy inverse problems},
\newblock in: \bibinfo{booktitle}{The Eleventh International Conference on Learning Representations}, \bibinfo{year}{2023}. \URLprefix \url{https://openreview.net/forum?id=OnD9zGAGT0k}.
%Type = Inproceedings
\bibitem[{Song et~al.(2023)Song, Vahdat, Mardani, and Kautz}]{song2023pseudoinverse}
\bibinfo{author}{J.~Song}, \bibinfo{author}{A.~Vahdat}, \bibinfo{author}{M.~Mardani}, \bibinfo{author}{J.~Kautz},
\newblock \bibinfo{title}{Pseudoinverse-guided diffusion models for inverse problems},
\newblock in: \bibinfo{booktitle}{International Conference on Learning Representations}, \bibinfo{year}{2023}.
%Type = Inproceedings
\bibitem[{Mardani et~al.(2024)Mardani, Song, Kautz, and Vahdat}]{mardani2023variational}
\bibinfo{author}{M.~Mardani}, \bibinfo{author}{J.~Song}, \bibinfo{author}{J.~Kautz}, \bibinfo{author}{A.~Vahdat},
\newblock \bibinfo{title}{A variational perspective on solving inverse problems with diffusion models},
\newblock in: \bibinfo{booktitle}{The Twelfth International Conference on Learning Representations}, \bibinfo{year}{2024}. \URLprefix \url{https://openreview.net/forum?id=1YO4EE3SPB}.
%Type = Article
\bibitem[{Karras et~al.(2022)Karras, Aittala, Aila, and Laine}]{karras2022elucidating}
\bibinfo{author}{T.~Karras}, \bibinfo{author}{M.~Aittala}, \bibinfo{author}{T.~Aila}, \bibinfo{author}{S.~Laine},
\newblock \bibinfo{title}{Elucidating the design space of diffusion-based generative models},
\newblock \bibinfo{journal}{Advances in neural information processing systems} \bibinfo{volume}{35} (\bibinfo{year}{2022}) \bibinfo{pages}{26565--26577}.
%Type = Article
\bibitem[{Lu et~al.(2022)Lu, Zhou, Bao, Chen, Li, and Zhu}]{lu2022dpm}
\bibinfo{author}{C.~Lu}, \bibinfo{author}{Y.~Zhou}, \bibinfo{author}{F.~Bao}, \bibinfo{author}{J.~Chen}, \bibinfo{author}{C.~Li}, \bibinfo{author}{J.~Zhu},
\newblock \bibinfo{title}{Dpm-solver: A fast ode solver for diffusion probabilistic model sampling in around 10 steps},
\newblock \bibinfo{journal}{Advances in Neural Information Processing Systems} \bibinfo{volume}{35} (\bibinfo{year}{2022}) \bibinfo{pages}{5775--5787}.
%Type = Inproceedings
\bibitem[{Liu et~al.(2022)Liu, Ren, Lin, and Zhao}]{liu2022pseudo}
\bibinfo{author}{L.~Liu}, \bibinfo{author}{Y.~Ren}, \bibinfo{author}{Z.~Lin}, \bibinfo{author}{Z.~Zhao},
\newblock \bibinfo{title}{Pseudo numerical methods for diffusion models on manifolds},
\newblock in: \bibinfo{booktitle}{International Conference on Learning Representations}, \bibinfo{year}{2022}. \URLprefix \url{https://openreview.net/forum?id=PlKWVd2yBkY}.
%Type = Inproceedings
\bibitem[{Salimans and Ho(2022)}]{salimans2022progressive}
\bibinfo{author}{T.~Salimans}, \bibinfo{author}{J.~Ho},
\newblock \bibinfo{title}{Progressive distillation for fast sampling of diffusion models},
\newblock in: \bibinfo{booktitle}{International Conference on Learning Representations}, \bibinfo{year}{2022}. \URLprefix \url{https://openreview.net/forum?id=TIdIXIpzhoI}.
%Type = Inproceedings
\bibitem[{Song et~al.(2023)Song, Dhariwal, Chen, and Sutskever}]{song2023consistencymodels}
\bibinfo{author}{Y.~Song}, \bibinfo{author}{P.~Dhariwal}, \bibinfo{author}{M.~Chen}, \bibinfo{author}{I.~Sutskever},
\newblock \bibinfo{title}{Consistency models},
\newblock in: \bibinfo{booktitle}{Proceedings of the 40th International Conference on Machine Learning}, ICML'23, \bibinfo{publisher}{JMLR.org}, \bibinfo{year}{2023}.
%Type = Inproceedings
\bibitem[{Chen et~al.(2022)Chen, Liu, and Theodorou}]{chen2022likelihood}
\bibinfo{author}{T.~Chen}, \bibinfo{author}{G.-H. Liu}, \bibinfo{author}{E.~Theodorou},
\newblock \bibinfo{title}{Likelihood training of schr\"odinger bridge using forward-backward {SDE}s theory},
\newblock in: \bibinfo{booktitle}{International Conference on Learning Representations}, \bibinfo{year}{2022}. \URLprefix \url{https://openreview.net/forum?id=nioAdKCEdXB}.
%Type = Inproceedings
\bibitem[{Chung et~al.(2024)Chung, Lee, and Ye}]{chung2023decomposed}
\bibinfo{author}{H.~Chung}, \bibinfo{author}{S.~Lee}, \bibinfo{author}{J.~C. Ye},
\newblock \bibinfo{title}{Decomposed diffusion sampler for accelerating large-scale inverse problems},
\newblock in: \bibinfo{booktitle}{The Twelfth International Conference on Learning Representations}, \bibinfo{year}{2024}. \URLprefix \url{https://openreview.net/forum?id=DsEhqQtfAG}.
%Type = Inproceedings
\bibitem[{Deng et~al.(2009)Deng, Dong, Socher, Li, Li, and Fei-Fei}]{deng2009imagenet}
\bibinfo{author}{J.~Deng}, \bibinfo{author}{W.~Dong}, \bibinfo{author}{R.~Socher}, \bibinfo{author}{L.-J. Li}, \bibinfo{author}{K.~Li}, \bibinfo{author}{L.~Fei-Fei},
\newblock \bibinfo{title}{Imagenet: A large-scale hierarchical image database},
\newblock in: \bibinfo{booktitle}{2009 IEEE conference on computer vision and pattern recognition}, \bibinfo{organization}{Ieee}, \bibinfo{year}{2009}, pp. \bibinfo{pages}{248--255}.
%Type = Inproceedings
\bibitem[{Karras et~al.(2018)Karras, Aila, Laine, and Lehtinen}]{karras2017progressive}
\bibinfo{author}{T.~Karras}, \bibinfo{author}{T.~Aila}, \bibinfo{author}{S.~Laine}, \bibinfo{author}{J.~Lehtinen},
\newblock \bibinfo{title}{Progressive growing of {GAN}s for improved quality, stability, and variation},
\newblock in: \bibinfo{booktitle}{International Conference on Learning Representations}, \bibinfo{year}{2018}. \URLprefix \url{https://openreview.net/forum?id=Hk99zCeAb}.
%Type = Inproceedings
\bibitem[{Karras et~al.(2019)Karras, Laine, and Aila}]{karras2019style}
\bibinfo{author}{T.~Karras}, \bibinfo{author}{S.~Laine}, \bibinfo{author}{T.~Aila},
\newblock \bibinfo{title}{A style-based generator architecture for generative adversarial networks},
\newblock in: \bibinfo{booktitle}{Proceedings of the IEEE/CVF conference on computer vision and pattern recognition}, \bibinfo{year}{2019}, pp. \bibinfo{pages}{4401--4410}.
%Type = Inproceedings
\bibitem[{Yu et~al.(2022)Yu, Zhang, Kang, Tang, Arnold, and Zhang}]{yu2022rplhr}
\bibinfo{author}{P.~Yu}, \bibinfo{author}{H.~Zhang}, \bibinfo{author}{H.~Kang}, \bibinfo{author}{W.~Tang}, \bibinfo{author}{C.~W. Arnold}, \bibinfo{author}{R.~Zhang},
\newblock \bibinfo{title}{Rplhr-ct dataset and transformer baseline for volumetric super-resolution from ct scans},
\newblock in: \bibinfo{booktitle}{International Conference on Medical Image Computing and Computer-Assisted Intervention}, \bibinfo{organization}{Springer}, \bibinfo{year}{2022}, pp. \bibinfo{pages}{344--353}.
%Type = Misc
\bibitem[{AAPM(2017)}]{Mayo_challenge}
\bibinfo{author}{AAPM}, \bibinfo{title}{Low dose ct grand challenge}, \bibinfo{howpublished}{[Online]}, \bibinfo{year}{2017}. \bibinfo{note}{Available: \url{http://www.aapm.org/GrandChallenge/LowDoseCT/}}.
%Type = Article
\bibitem[{Heusel et~al.(2017)Heusel, Ramsauer, Unterthiner, Nessler, and Hochreiter}]{heusel2017gans}
\bibinfo{author}{M.~Heusel}, \bibinfo{author}{H.~Ramsauer}, \bibinfo{author}{T.~Unterthiner}, \bibinfo{author}{B.~Nessler}, \bibinfo{author}{S.~Hochreiter},
\newblock \bibinfo{title}{Gans trained by a two time-scale update rule converge to a local nash equilibrium},
\newblock \bibinfo{journal}{Advances in neural information processing systems} \bibinfo{volume}{30} (\bibinfo{year}{2017}).
%Type = Inproceedings
\bibitem[{Zhang et~al.(2018)Zhang, Isola, Efros, Shechtman, and Wang}]{zhang2018perceptual}
\bibinfo{author}{R.~Zhang}, \bibinfo{author}{P.~Isola}, \bibinfo{author}{A.~A. Efros}, \bibinfo{author}{E.~Shechtman}, \bibinfo{author}{O.~Wang},
\newblock \bibinfo{title}{The unreasonable effectiveness of deep features as a perceptual metric},
\newblock in: \bibinfo{booktitle}{CVPR}, \bibinfo{year}{2018}.
%Type = Article
\bibitem[{Bishop(2006)}]{bishop2006pattern}
\bibinfo{author}{C.~M. Bishop},
\newblock \bibinfo{title}{Pattern recognition and machine learning},
\newblock \bibinfo{journal}{Springer google schola} \bibinfo{volume}{2} (\bibinfo{year}{2006}) \bibinfo{pages}{1122--1128}.
%Type = Inproceedings
\bibitem[{Jabri et~al.(2023)Jabri, Fleet, and Chen}]{jabri2023scalable}
\bibinfo{author}{A.~Jabri}, \bibinfo{author}{D.~J. Fleet}, \bibinfo{author}{T.~Chen},
\newblock \bibinfo{title}{Scalable adaptive computation for iterative generation},
\newblock in: \bibinfo{booktitle}{Proceedings of the 40th International Conference on Machine Learning}, \bibinfo{year}{2023}, pp. \bibinfo{pages}{14569--14589}.

\end{thebibliography}

\end{document}